\documentclass[aps,amsmath,amssymb,preprintnumbers,nofootinbib,11pt]{revtex4}
\usepackage[utf8]{inputenc} %utf8
\usepackage{graphicx}
\graphicspath{{./figures/}}
\usepackage{url}
\usepackage[bookmarks, pagebackref=false]{hyperref}
\usepackage[usenames,dvipsnames]{xcolor}

\usepackage{makecell}

\definecolor{orange}{cmyk}{0,0.5,1,0}
\definecolor{rossoCP3}{cmyk}{0,.88,.77,.40}
\definecolor{graa}{rgb}{0.8,0.8,0.8}
\definecolor{blaa}{rgb}{0.2,0.2,0.6}
	\hypersetup{
			colorlinks,
			bookmarksopen,
			bookmarksnumbered,
			citecolor=blaa, 		%color of links to bibliography
			linkcolor=rossoCP3,	%color of internal links
			urlcolor=rossoCP3,			%color of external links
			}
\usepackage{amsthm}
\usepackage{bm}% bold math
\usepackage{bbm}
\usepackage{pxfonts}

\usepackage{amsmath,amssymb,amsfonts}
\usepackage{color}
\usepackage{float}
\usepackage{hyperref}
\usepackage[Symbolsmallscale]{upgreek}
\usepackage{amsmath}
\usepackage{amsfonts}
\usepackage{amssymb,dsfont}
\usepackage{graphicx}
\usepackage{amssymb}
\usepackage[vcentermath]{youngtab}
\usepackage[all]{xy}
\usepackage{pstricks}
\usepackage{dsfont}%
\usepackage{bbold}
\usepackage[normalem]{ulem}

\usepackage{placeins}
\usepackage{xspace}
\usepackage{cancel} % to make the barred text notation

\usepackage{slashed}
\usepackage{natbib}

\usepackage{feynmf}

\usepackage{braket}

\newcommand{\beq}{\begin{eqnarray}}
\newcommand{\eeq}{\end{eqnarray}}

\newcommand{\bmp}{\noindent\begin{minipage}{16cm}}
\newcommand{\emp}{\end{minipage}\vskip 7mm} % 7mm untightened

\newcommand   \cO {\mathcal{O}}

\newcommand   \cA {\mathcal{A}}

\newcommand{\eps}  {\epsilon}

\def\lsim{\mathrel{\rlap{\lower4pt\hbox{\hskip1pt$\sim$}}
    \raise1pt\hbox{$<$}}}                % less than or approx. symbol
\def\gsim{\mathrel{\rlap{\lower4pt\hbox{\hskip1pt$\sim$}}
    \raise1pt\hbox{$>$}}}                % greater than or approx. symbol

\begin{document}
%%%%%%%%%%%%%%%%%%%%%%%%%%%%%%%%%%%%%%%%%%%%%%%%%%%%%%%%%%%%%%%%%%%%%%%%%%%

\title{More on the cubic versus quartic interaction equivalence in the O(N) model }
\author{Oleg {\sc Antipin}$^{\color{rossoCP3}{\clubsuit}}$}
\email{oantipin@irb.hr}
\author{Jahmall {\sc Bersini}
$^{\color{rossoCP3}{\clubsuit}}$}
\email{jbersini@irb.hr}
\author{Francesco {\sc Sannino} $^{\color{rossoCP3} {\diamondsuit},\color{rossoCP3}{{\varheartsuit},\heartsuit}}$}
\email{sannino@cp3.dias.sdu.dk}
\author{Zhi-Wei Wang $^{\color{rossoCP3}{\diamondsuit}}$}
\email{wang@cp3.sdu.dk}
\author{Chen Zhang $^{\color{rossoCP3}{\spadesuit}}$}
\email{chen.zhang@fi.infn.it}
\affiliation{{ $^{\color{rossoCP3}{\clubsuit}}$ Rudjer Boskovic Institute, Division of Theoretical Physics, Bijeni\v cka 54, 10000 Zagreb, Croatia}\\{ $^{\color{rossoCP3}{\diamondsuit}}$\color{rossoCP3} {CP}$^{ \bf 3}${-Origins}} \& the Danish Institute for Advanced Study {\color{rossoCP3}\rm{Danish IAS}},  University of Southern Denmark, Campusvej 55, DK-5230 Odense M, Denmark. \\
\mbox{ $^{\color{rossoCP3}{\varheartsuit}}$ Scuola Superiore Meridionale, \\ Largo S. Marcellino, 10, 80138 Napoli, Italy}
\mbox{  $^{\color{rossoCP3}{\heartsuit}}$ Dipartimento di Fisica, E. Pancini, Universit\'a di Napoli, Federico II, INFN sezione di Napoli} \\ \mbox{Complesso Universitario di Monte S. Angelo Edificio 6, via Cintia, 80126 Napoli, Italy.}\\
{$^{\color{rossoCP3}{\spadesuit}}$ INFN Sezione di Firenze, Via G. Sansone 1, I-50019 Sesto Fiorentino, Italy}}

\begin{abstract}
{We compute the scaling dimensions of a family of fixed-charge operators at the infrared fixed point of the $O(N)$ model featuring cubic interactions in $d=6-\eps$ for arbitrary $N$ to leading and subleading order in the charge but to all orders in the couplings. The results are used to  analyze the conjectured equivalence with the $O(N)$  model displaying quartic interactions at its ultraviolet fixed point. This is performed  by comparing the cubic model scaling dimensions against the known large $N$ results for the  quartic model and demonstrating that they match.
 Our results reinforce the conjectured equivalence and further provide novel information on the finite $N$ physics stemming from the computations in the cubic model just below 6 dimensions.}
  \\~\\
{\footnotesize  \it Preprint: RBI-ThPhys-2021-27}

\end{abstract}

\maketitle

\small
\newpage

\section{Introduction}
The $O(N)$ model with quartic interactions has a long history with applications ranging from condensed matter to high energy physics \cite{HKleinertFrohlinde, Pelissetto:2000ek, Zhang, YANG, deGennes:1972zz, Pisarski:1983ms} (e.g.~the Standard Model Higgs) including string theory \cite{Klebanov:2002ja, Giombi:2009wh}. It has been investigated in several space-time dimensions and its dynamics discussed within and beyond perturbation theory and/or in different thermodynamic regimes of temperature and matter density. Despite the relevance and the energies devoted its full dynamics remains unknown. For example, consider the theory just below six dimensions. It was Parisi \cite{Parisi:1975im} to show that in the large $N$ expansion and for $4<d<6$ an ultraviolet (UV) fixed point (FP) emerges in the $O(N)$ model with quartic couplings, rendering the theory non-perturbatively renormalizable in the $1/N$ expansion. At the same time, the finite $N$ dynamics is not yet solved. Is it possible to make progress beyond the large $N$ limit? In order to positively answer this question and make a dent in this direction, we will consider a slight detour. The latter uses an alternative model, known as the cubic model, according to which rather than having quartic interactions the theory features an $O(N)$ singlet field interacting with the $O(N)$ fields via a cubic operator \cite{Fei:2014yja}. The relation between the two models is expected to hold in between four and six dimensions
% . The reason for such a relation may be partially traced back to the fact that a singlet scalar field can be  introduced to reduce the order of the quartic interactions. In this case, it starts out as an auxiliary field, meaning it does not possess a kinetic term   but it naturally displays cubic interactions with the $O(N)$.  The kinetic term and further cubic interactions appear because of the dynamics.
% Whatever the origin of the cubic model one expects the conjectured equivalence with the quartic $O(N)$ model  to hold for $4 < d < 6$
because both theories are simultaneously critical in this range.

In fact, in \cite{Fei:2014yja}, the authors proposed a dual description of the UV FP of the quartic theory in $4<d<6$ in terms of the infrared (IR) FP of the cubic model.
% $O(N)$ model coupled to an extra singlet scalar field through a cubic interaction.
For the cubic theory, one can show that it has upper critical dimension $d=6$ and thus it is usually investigated in $d=6-\eps$ via the $\eps$-expansion.  Within this expansion one observes a critical value of $N$ above which the model features an IR FP.  The one-loop guesstimate obtained pushing $\eps$ to one gives  $N_{\rm crit} = 1038$  \cite{Fei:2014yja} that should be confronted with the four- and five-loop results  resummed results \cite{Gracey:2015tta, Kompaniets:2021hwg} that place it at $N_{\rm crit} \approx 400$.

The equivalence between the quartic and the cubic critical theories has been supported by a series of investigations regarding the scaling dimensions at the cubic FP of various operators \cite{Gracey:2015tta, Kompaniets:2021hwg, Fei:2014xta, Fei:2014yja, Arias-Tamargo:2020fow}, and few OPE coefficients \cite{Goykhman:2019kcj}, which strikingly match their counterparts in the critical quartic theory.
Since the cubic and quartic models are usually investigated via the $\eps$ and $1/N$ expansion, respectively, the comparison is usually performed at the level of the terms which appear in both expansion schemes at the considered order of expansion. In Table \ref{tab:summary} (at the end of Section ~\ref{vivi}), we  summarize, together along with the new results reported in the present paper, the checks of the dual picture appeared in the literature so far.

On the other hand, even if the critical $O(N)$ theory is well-defined and unitary order by order in the $1/N$ expansion, the quartic theory cannot exhibit a stable FP \cite{Aizenman:1981du, Rosten:2008ts, Percacci:2014tfa} because it occurs at negative values of the quartic coupling. In the cubic theory, the instability of the quartic FP is manifested by the fact that the cubic potential is unbounded from below. As shown in \cite{Giombi:2019upv}, the instability is realized by instantonic effects which mediate the tunnelling from the vacuum at $0$-values of the fields to large negative values of the singlet scalar field and give rise to complex CFT data. Remarkably, in \cite{Giombi:2019upv}, the non-perturbative instantonic contribution to the CFT data has been computed in both theories finding agreement between the two pictures. This contribution is exponentially suppressed at large $N$. As a result, when $N$ is large enough, the imaginary part of the CFT data is negligible and the UV FP of the quartic theory can be studied via the conformal bootstrap as done in \cite{Chester:2014gqa, Bae:2014hia, Li:2016wdp}, where the authors found a region in the parameter space where the CFT data are in good agreement with the results from the $1/N$ expansion. The instability of the cubic FP at large $N$ has been confirmed by functional studies \cite{Kamikado:2016dvw, Eichhorn:2016hdi}.

Here we use large-charge semiclassical methods
%\footnote{It may be more precisely called semiclassical fixed-charge expansion.}
\cite{Arias-Tamargo:2019xld,
Antipin:2021akb, Antipin:2020rdw, Antipin:2020abu, Dondi:2021buw, Badel:2019oxl, Gaume:2020bmp, Alvarez-Gaume:2019biu, Alvarez-Gaume:2016vff, Giombi:2020enj, Hellerman:2015nra, Cuomo:2021ygt, Orlando:2020yii, Orlando:2019skh, Orlando:2019hte, Hellerman:2017sur, delaFuente:2018qwv, Banerjee:2017fcx, Jack:2020wvs, Jack:2021ypd}
  to further test the equivalence between the two CFTs.  We compute the scaling dimensions of a whole family of fixed-charge operators for the cubic model in $d=6-\eps$ and compare the results with the existing $O(N)$ literature. In particular, our semiclassical expansion resums an infinite series of Feynman diagrams and allows comparing terms at arbitrarily high perturbative orders providing interesting insights of the dual picture.

The first test of the equivalence of the large-charge sector has been carried out in \cite{Arias-Tamargo:2020fow}, where the authors calculated the scaling dimension of traceless symmetric $O(N)$ operators in $d=6-\eps$ dimensions at the leading order in both the charge and $\eps$, obtaining the same result in both models\footnote{This result is obtained from the exponentiation of the diagrams with the leading charge-scaling at every loop order.}.

 We summarize below the highlights of our work:

\begin{itemize}
\item[1.]We compute the scaling dimensions of a family of fixed-charge operators at the infrared fixed point of the $O(N)$ model featuring cubic interactions in $d=6-\eps$ to leading (LO) and next-to-leading (NLO) order in the fixed-charge expansion for arbitrary $N$ and to all orders in the cubic model couplings.

\item[2.] We use the results above to  analyze the envisioned equivalence with the critical $O(N)$  model with quartic interactions by comparing them against the known large $N$ results for the  quartic  model. The results are summarized in Table \ref{tab:summary} (at the end of Section \ref{vivi}).

\item[3.] Once established that our results further support the equivalence of the two model descriptions of the critical dynamics we provide novel results on the finite $N$ physics stemming from the cubic computation near $d=6$ dimensions.

\item[4.] We analyze the onset of complex CFT dynamics of the large charge sectors of the two models by determining the critical charge above which the anomalous dimensions become complex.
\end{itemize}

The paper is organised as follows. In \autoref{I} we review large-$N$ results on the scaling dimension of traceless symmetric $O(N)$-operators in the quartic model. Then, in \autoref{vivi}, we semiclassically compute the same quantities in the cubic theory and we compare them with the quartic model's results mentioned above. In \autoref{complex}, we study the emergence of complex CFT data above a critical value of the charge, which we compute in the $\eps$-expansion. We offer our conclusions in \autoref{discuss}. Appendix \ref{rino}, contains details on the $1$-loop renormalization of the semiclassical expansion in the cubic model.

\section{The $O(N)$ quartic model in $4<d<6$: traceless symmetric operators} \label{I}

In this section, we collect relevant results in the literature about the scaling dimensions of operators transforming in the traceless symmetric representations of the $O(N)$ group at the UV FP of the $O(N)$ quartic model in $4<d<6$ dimensions. In the next section, we will semiclassically compute the same quantities at the IR FP of the $O(N)$ cubic model in $6-\epsilon$ dimensions and compare the results. Their equality will provide a stringent test of the proposed equivalence between the two FPs, at least at large $N$ and near six dimensions. The Euclidean action of the $O(N)$ quartic model in $d$ dimensions is written as
\begin{equation} \label{action}
    \mathcal{S}=\int d^d x \left(\frac{(\partial \phi_a)^2}{2}+\frac{(4\pi)^2 g_0}{4!}(\phi_a\phi_a)^2\right) \ .
\end{equation}
Here $\phi_a, \ a=1,2,...,N$ are a set of $N$ real scalar fields which collectively transform as an $O(N)$ vector. In the above expression, the summation over $a$ from $1$ to $N$ is understood. In what follows, we will be interested in operators transforming as traceless fully symmetric tensors of $O(N)$, and that can be expressed as
\begin{equation}
    T_Q =  t_Q^{a_1...a_Q}(\phi_a) \ ,
\label{pmldef}
\end{equation}
where $t_Q^{a_1...a_Q}(\phi_a)$ is a homogeneous polynomial of degree $Q$ in the $\phi_a$'s that is fully symmetric in the indices and traceless (i.e.~contraction of any two indices gives zero).
The explicit form of the first three $t_Q^{a_1...a_Q}$ polynomials reads
\begin{align}
t^{a}_{1}(\phi) & = \phi^a \ , \quad \qquad
t^{ab}_{2}(\phi)= \phi^a \phi^b - {1\over N} \delta^{ab} \phi^2  \ , \quad \qquad
t^{abc}_3(\phi) = \phi^a \phi^b \phi^c - {\phi^2\over N + 2} \left(\phi^a
  \delta^{bc} + \phi^b \delta^{ac} + \phi^c \delta^{ab}\right) \ ,
\label{spin4}
\end{align}
where $\phi^2\equiv \sum_{a=1}^{a=N} \phi_a\phi_a$.
% \tad{Physically, the $T_{Q}$'s represent anisotropic perturbations in $O(N)$-invariant systems and their scaling dimensions control how the perturbations to influence the critical behavior. These operators are of interested for a variety of three-dimensional systems, such as density-wave systems \cite{Brock-etal-86,ABBL-86}, magnets with a cubic crystal structure \cite{Aharony-76, Antipin:2019vdg}, and superconductors \cite{Zhang, Calabrese:2002bm}.}
The classical scaling dimension of $T_Q$ is $Q\,(d/2-1)$. Furthermore, $Q$ can be identified with its total charge as follows.
For simplicity, suppose $N$ is even, then we can define $N/2$ Cartan charges as
\begin{align} \label{cartancharges}
\mathcal{Q}^{(j)}=\int d^{d-1}x(\phi_{2j-1}\partial^0\phi_{2j}-\phi_{2j}\partial^0\phi_{2j-1}), \ j=1,2,...,N/2 \ ,
\end{align}
and fix their values as $\mathcal{Q}^{(j)} = Q^{(j)}$ with $\left\{Q^{(j)}\right\}$ a set of constants.
It is then possible to show that all operators that have the same value of $Q = \sum_{j=1}^{j=N/2}|Q^{(j)}|$ and have the minimal classical scaling dimension span the same space as the operator $T_Q$~\cite{Antipin:2020abu}. Then, in the perturbative regime, i.e. when the anomalous dimensions are small, the $T_Q$ are the lowest-lying $O(N)$-operators with total charge $Q$. As we shall see in the next section, this property will be relevant for our semiclassical computations.

At the fixed point, the scaling dimension of the $T_Q$ operators (denoted $\Delta_Q$) can be computed non-perturbatively via Monte Carlo simulations \cite{Hasenbusch_2011} and the conformal bootstrap method \cite{Chester:2020iyt}. Furthermore, $\Delta_Q$ is usually computed perturbatively in the $\epsilon$-expansion in $d = 4 - \eps $ dimensions~\cite{Kehrein:1995ia, Kompaniets:2019zes, Calabrese:2002bm} and in the large $N$ expansion
~\cite{Lang:1992zw,Derkachov:1997ch}.
Nice expositions of these conventional approaches are available in the literature, e.g. ref.~\cite{HKleinertFrohlinde} for the $\epsilon$-expansion, and ref.~\cite{Moshe:2003xn} for the large $N$ expansion. While most of the literature investigated $\Delta_Q$ in the range $2<d<4$, which is relevant for condensed matter experiments, the $1/N$ expansion is well-defined and unitary also in $4<d<6$. The upper limit $d<6$ is set by observing that the scaling dimension at the UV FP of the $\phi^2$ operators is $2+\cO\left(\frac{1}{N}\right)$ in any $d$, and thus it violates the unitarity bound ($d/2-1$) when $d>6$ \cite{Fei:2014yja}.

The large $N$ expansion is generated by performing a Hubbard-Stratonovich transformation, turning the Lagrangian \eqref{action} into
\begin{equation}
\mathcal{S} = \int d^d x \left(\frac{1}{2}(\partial \phi_a)^2+\frac{1}{2}\sigma \phi_a\phi_a-\frac{3\sigma^2}{2 g_0 (4 \pi)^2}\right)\ ,
\label{hubbard}
\end{equation}
where we introduced an auxiliary field $\sigma$ that can be integrated out via its equation of motion (EOM) $\sigma = \frac{(4 \pi)^2 g_0}{6} \phi_a \phi_a$ to come back to the original action \eqref{action}. At the critical point, it is possible to neglect the last term and one is left with the following action \cite{Giombi:2016ejx}
\begin{equation}
\mathcal{S}_{\rm crit} = \int d^dx \left(\frac{1}{2}(\partial \phi_a)^2+\frac{1}{2}\sigma \phi_a\phi_a\right) \ .
\label{S-crit}
\end{equation}
The $1/N$ expansion is then generated by integrating out the fields $\phi_a$, which appear quadratic in $\mathcal{S}_{\rm crit}$. $\Delta_Q$ has been computed in the $1/N$ expansion for arbitrary $Q$ and $d$ to order $\cO\left(\frac{1}{N^2} \right)$ \cite{Lang:1992zw, Derkachov:1997ch}
\begin{align} \label{totaldiagr}
\Delta_Q &= \left(\frac{d}{2}-1\right)Q -\frac{1}{N}\left[\frac{2^{d-3} d \sin \left(\frac{\pi  d}{2}\right) \Gamma \left(\frac{d-1}{2}\right)}{\pi ^{3/2} \Gamma \left(\frac{d}{2}+1\right)}\left(Q(Q-2)+\frac{4Q}{d}\right)\right] + \frac{1}{N^2} \Bigg[\eta Q-\frac{8 (Q-1) Q \sin ^2\left(\frac{\pi d}{2}\right) \Gamma (d-2)^2}{3 \pi ^2 (d-2)^3 d \Gamma \left(\frac{d}{2}-1\right)^4}  \nonumber \\ &  \times \left(-12 ((d-3) d+4) (d-2) \left(H_{d-3}+\pi  \cot
   \left(\frac{\pi  d}{2}\right)\right)+d (d-2)^2 ( Q-2) \left(\pi
   ^2-6 \psi ^{(1)}\left(\frac{d}{2}\right)\right)+12 (d-3)
   d\right)\Bigg] +\cO\left(\frac{1}{N^3}\right)\ ,
\end{align}
where $H$ denotes the harmonic numbers and $\eta$ is the coefficient of the $\left(\frac{1}{N^2} \right)$ term in the scaling dimension of $\phi_a$. The latter has been computed to order $\cO \left(\frac{1}{N^3} \right)$ in \cite{Vasiliev:1992wr}.
In $d=6-\eps$, we have $\eta = 44 \eps - \frac{835}{6} \eps^2+ \cO\left( \eps^3\right)$, and the above becomes %\tad{Chen: Is it possible to highlight the following equation?}
\begin{align} \label{1loopfull}
   \Delta_Q &= 2 Q - \frac{\eps}{2}Q+ \frac{1}{N} \left[ \left( - 3Q^2+ 4 Q  \right)\eps +\left( \frac{7  }{4 }Q^2 -\frac{8}{3}Q \right)\epsilon^2+\cO \left(\eps^3 \right)  \right] \nonumber \\ &+ \frac{1}{N^2}\left[ \left( - 132Q^2+ 176 Q  \right)\eps    -\left( 45 Q^3 -\frac{857}{2}Q^2 +\frac{1568}{3}Q \right)\epsilon^2+\cO \left(\eps^3 \right)  \right] + \cO\left(\frac{1}{N^3}\right) \ .
\end{align}

Recently, these operators have been studied in the double scaling limit $Q \to \infty$, $N \to \infty$ with fixed 't Hooft-like coupling $J \equiv Q/N$ in $d=3$  \cite{Alvarez-Gaume:2019biu} and $2<d<6$ \cite{Giombi:2020enj}. In this limit, the scaling dimension of $T_Q$ takes the form
\begin{equation}\label{expansion1}
\Delta_Q =\sum_{k=-1} \frac{1}{N^k}\Tilde{\Delta}_k (J) \ ,
\end{equation}
corresponding to a semiclassical expansion around the classical solution of the theory at fixed-charge. The LO $\Tilde{\Delta}_{-1}$ is the classical term of the saddle-point expansion and has been computed in \cite{Giombi:2020enj}
\begin{equation}
\begin{aligned}
\Tilde{\Delta}_{-1} =  \left[f(c_{\sigma})+J \sqrt{ \left( \frac{d}{2}-1 \right)^2 + c_{\sigma}} \right]_{c_{\sigma}=c_{\sigma}(J)} \ ,
\end{aligned}
\label{Delta-final}
\end{equation}
where $c_{\sigma}(J)$ solves
\begin{equation}
    \frac{d}{dc_{\sigma}} \left[f(c_{\sigma}) + J \sqrt{ \left( \frac{d}{2}-1 \right)^2 + c_{\sigma}}  \right] = 0\ ,
    \label{saddle-csol-extra}
\end{equation}
and $f(c_{\sigma})$ is given by
\begin{equation}
f(c_{\sigma}) = - \frac{c_{\sigma}}{d-2} \int_0^{\infty} dt \, \frac{J_2 \left( \sqrt{c_{\sigma}} t \right)}{t(2 \cosh t - 2)^{\frac{d}{2}-1}}\ ,
\label{fc}
\end{equation}
with $J_2$ the Bessel function of the first kind. The small $J$ expansion of $\Tilde{\Delta}_{-1}$ reads \cite{Giombi:2020enj}
\begin{equation}
\Tilde{\Delta}_{-1} =  \left(\frac{d}{2}-1\right)J+h_2(d)J^2+h_3(d)J^3+\ldots \ ,
\label{Delta-smallj}
\end{equation}
where
\begin{equation}
\begin{aligned}
&h_2(d) =-\frac{2^{d-3} d \sin \left(\frac{\pi  d}{2}\right) \Gamma \left(\frac{d-1}{2}\right)}{\pi ^{3/2} \Gamma \left(\frac{d}{2}+1\right)} \ , \qquad h_3(d)= -\frac{(d-2) d^2 \Gamma (d-2)^2 \left(\pi ^2-6 \psi ^{(1)}\left(\frac{d}{2}\right)\right)}{6 \Gamma \left(2-\frac{d}{2}\right)^2 \Gamma \left(\frac{d}{2}-1\right)^4 \Gamma \left(\frac{d}{2}+1\right)^2} \, \qquad h_4(d)= \dots
\label{h2h3}
\end{aligned}
\end{equation}
$\Tilde{\Delta}_{-1}$ resums an infinite number of terms of the conventional large $N$ expansion. Specifically, it resums all the terms with the leading $Q$-scaling at every $1/N$ order\footnote{In fact it can be checked that $h_2(d)$ and $h_3(d)$ agree with the diagrammatic result \eqref{totaldiagr}.}. This fact will allow us to probe the equivalence of quartic and cubic theories by comparing terms up to arbitrarily high orders in the $1/N$ expansion by performing a similar computation in the cubic theory.
In $d=6-\epsilon$, $\Tilde{\Delta}_{-1}$ can be expanded as
\begin{align} \label{compa}
N  \Tilde{\Delta}_{-1} &= 2 Q -\frac{\eps}{2}Q+ Q \sum_j  \left(\frac{Q}{N}\right)^j\left(\alpha_j \eps^j + \beta_j \eps^{j+1}+ \gamma_j \eps^{j+2} + \dots\right) \ .
\end{align}
For later comparison with the cubic model, we list below the values of the first $\alpha_j$ and $\beta_j$ coefficients
%\tad{Chen: Is it possible to highlight Eq.~\eqref{alfas} and Eq.~\eqref{betas}?}
\begin{align} \label{alfas}
& \alpha_1 = -3 \ , \qquad \alpha_2 = -45 \ , \qquad \alpha_3 = -1350 \ , \qquad \alpha_4 = -\frac{213597}{4}  \ , \qquad \alpha_5 = - 2457216 \nonumber \\ & \alpha_6 = - \frac{995773905}{8}  \ , \qquad \alpha_7 = - 6739459200  \ , \qquad \alpha_8 = - \frac{24526111620285}{64} \ .
\end{align}
\begin{align} \label{betas}
& \beta_1 = \frac{7}{4} \ , \qquad \beta_2 =\frac{3}{4} (48 \zeta (3)+31) \ , \qquad \beta_3 =\frac{27}{2} (128 \zeta (3)+40 \zeta (5)+41) \ , \nonumber \\ & \beta_4 =\frac{81}{16} (18208 \zeta (3)+7168 \zeta (5)+1792 \zeta (7)+3117)  \ , \quad \beta_5 =648 (8202 \zeta (3)+3510 \zeta (5)+1218 \zeta (7)+252 \zeta (9)+677)\nonumber \\ &  \beta_6 = \frac{2187}{32} (4727888 \zeta (3)+2109440 \zeta (5)+836864 \zeta
   (7)+256000 \zeta (9)+45056 \zeta (11)+110211) \ .
\end{align}

It is interesting to consider also the large $J$ expansion of $\Tilde{\Delta}_{-1}$, which reads
\begin{equation}
    \Delta_Q = N J^{\frac{d}{d-1}}\left(\delta_0 +  \delta_1 J^{\frac{-2}{d-1}}+  \delta_2 J^{\frac{-4}{d-1}}+ \ldots \right)  \ ,
\label{del-largej}
\end{equation}
where \cite{Giombi:2020enj}
\begin{align}
    \delta_0 =\left(1-\frac{1}{d}\right)C_0^{\frac{1}{2}} \ , \qquad \delta_1 =\frac{(d-1)(d-2)}{12}C_0^{-\frac{1}{2}} \ , \qquad \delta_2 =- \frac{(d-1)(d-2)^2(3d-2)}{1440}C_0^{-\frac{3}{2}}  \ ,
\end{align}
with
\begin{align}
    C_0 = \left(-\frac{2^d}{\pi d} \sin \left(\frac{\pi d}{2}\right)\Gamma \left(\frac{d}{2}\right) \left(1+\frac{ d}{2}\right) \right)^{\frac{2}{d-1}} \ .
\end{align}
Eq.\eqref{del-largej} agrees with the general form of the large-charge expansion in $O(N)$ symmetric models \cite{Hellerman:2015nra, Gaume:2020bmp}
\begin{align}
\Delta_{Q}= Q^{\frac{d}{d-1}}\left[\alpha_{1}+ \alpha_{2} Q^{\frac{-2}{d-1}}+\alpha_3 Q^{\frac{-4}{d-1}}+\ldots\right] +Q^0\left[\beta_0+ \beta_{1} Q^{\frac{-2}{d-1}}+\ldots\right] + \cO\left( Q^{-\frac{d}{d-1}}\right)  \ ,
\label{largecharge}
\end{align}
which has been predicted using effective field theory methods which do not rely on the presence of other expansion parameters besides $Q$. Since $C_0$ is complex in $4<d<6$, the scaling dimensions in the large $J$ expansion are complex as well. It can be shown that the imaginary part in $\Delta_Q$ arises at a critical value of $J$,  above which there are no real solutions to the saddle-point equations \cite{Giombi:2020enj}. $J_c$ depends non-trivially on $d$ and has been estimated numerically in \cite{Giombi:2020enj} in the whole range $4<d<6$. Accordingly, we shall see that also the cubic FP exhibits the presence of a critical charge. Assuming the validity of the dual description, in Sec.\ref{complex} we will compute $J_c$ analytically in the $\eps$ expansion in both $d=6-\eps$ and $d=4+\eps$. In $d= 6-\eps$ dimensions the large $J$ expansion of $\Tilde{\Delta}_{-1}$ reads %\tad{Chen: Is it possible to highlight the following equation?}
\begin{equation} \label{leadinglarge}
   N \Tilde{\Delta}_{-1} = -e^{\pm i4\pi/5}\frac{5N}{3}(2\eps)^{1/5} J^{6/5}(1+ \cO(\eps)) + e^{\pm i\pi/5}\frac{5N}{6}(2\eps)^{-1/5} J^{4/5}(1+ \cO(\eps)) - e^{\pm 3 i\pi/5}\frac{N}{9}(2\eps)^{-3/5} J^{2/5}(1+ \cO(\eps)) +\cO \left(J^0 \right) \ .
\end{equation}

\section{The semiclassical expansion in the cubic model} \label{vivi}

As mentioned, an alternative description of the critical quartic $O(N)$ model in $4<d<6$, is given by the infrared FP of a theory with $N+1$ fields, $O(N)$ symmetry, and Lagrangian
 \begin{equation} \label{cubicmodel}
    \mathcal{L} =
  \frac{1}{2} (\partial  \phi_a)^2+ \frac{1}{2} (\partial \eta)^2 + \frac{g_0}{2}\eta (\phi_a)^2 + \frac{h_0}{6} \eta^3 \ ,
\end{equation}
where $\phi_a$ is again an $O(N)$ vector. This model is usually studied near its upper critical dimension, $d=6$, where the infrared dynamics becomes free. In $d=6-\eps$, the $1$-loop beta functions of the model read
\begin{equation}
\beta_g = -\frac{\epsilon}{2}g+\frac{(N-8)g^3-12g^2 h + g h^2}{12(4\pi)^3}\ , \qquad \beta_h =-\frac{\epsilon}{2}h+ \frac{-4Ng^3 +Ng^2 h -3h^3}{4(4\pi)^3}\ ,
\label{beta1}
\end{equation}
% and at LO in large $N$ simplify as
% \begin{align} \label{betaN}
% \beta_g = -\frac{\epsilon}{2}g + \frac{Ng^3}{12(4\pi)^3} \ , \qquad \beta_h = -\frac{\epsilon}{2}h + \frac{-4Ng^3 +Ng^2 h}{4(4\pi)^3}\ .
% \end{align}

At large enough $N$, the model features an IR FP at real values of the two couplings which, at the one-loop level, read
\begin{align}
g^* &\equiv \sqrt{\frac{6\eps(4\pi)^3}{N}}\left( 1 + \frac{22}{N}+\frac{726}{N^2}-\frac{326180}{N^3}  +... +  \cO\left(\eps \right)  \right) \ , \nonumber \\  h^* &\equiv 6 \sqrt{\frac{6\eps(4\pi)^3}{N}}\left(1 + \frac{162}{N}+\frac{68766}{N^2}+\frac{41224420}{N^3}   +... +  \cO\left(\eps \right)  \right) \ .
\label{gtoxy}
\end{align}
The $5$-loop beta functions have been derived in \cite{Kompaniets:2021hwg}. On can check that at the leading order in $1/N$, the FP couplings are exact at order $\sqrt{\epsilon}$.

\subsection{Leading order}

We now proceed by computing the $\Delta_Q$ to NLO in the semiclassical large-charge expansion, i.e in the double scaling limit $\eps \to 0$, $Q \to \infty$ with $\mathcal{A}\equiv Q \eps$ fixed.
We start by introducing $N/2$ complex fields as
\begin{equation}
\varphi_j=\frac{1}{\sqrt{2}}( \phi_{2 j - 1} + i \phi_{2j})
% = \frac{1}{\sqrt{2}}\rho_j e^{i \chi_j}
\ ,  \qquad j=1, \dots, N/2 \ ,
\end{equation}
and mapping the theory to the cylinder $\mathbb{R}^d \to \mathbb{R} \times S^{d-1}$ \cite{Rychkov:2016iqz}. Considering polar coordinates $(r, \Omega_{d-1})$ for $\mathbb{R}^d$, the map reads
\begin{equation}
  (r, \Omega_{d-1}) \to  (\tau, \Omega_{d-1}) \ ,  \qquad   r= R e^{\tau/R} \ ,
\end{equation}
 with $R$ the radius of $S^{d-1}$. The cylinder Lagrangian
 \begin{align} \label{modelcub}
    \mathcal{L}_{\rm cyl} &=
 \partial \varphi_j^* \partial \varphi_j  + \frac{1}{2} \partial \eta \partial \eta + g_0 \eta (\varphi_j^* \varphi_j) + \frac{h_0}{6} \eta^3 + \frac{m^2}{2}\eta^2+ m^2 \varphi_j^* \varphi_j \ ,
\end{align}
contains mass terms stemming from the conformal coupling of the fields to the Ricci scalar of the cylinder \cite{Brown:1980qq}. The mass reads $m=\frac{d-2}{2R}$. According to the state-operator correspondence (e.g.~\cite{Rychkov:2016iqz}), the action of an operator $\tau = -\infty$ creates a state on the cylinder with the same quantum numbers and with energy related to its scaling dimension by
\begin{equation}
\label{eedelta}
 E = \frac{\Delta}{R}  \ .
\end{equation}
As anticipated, $T_Q$ is the lowest-lying operator with total charge $Q$, and, as a consequence, we can compute $\Delta_Q$ by considering the expectation value of the evolution operator $e^{-H T}$ (with $H$ the Hamiltonian and $T = \tau_f - \tau_i$) in an arbitrary state $\ket{Q}$ with fixed (total) charge $Q$ and taking the limit $T \to \infty$ in order to project out the ground state from it. That is
\begin{equation} \label{evolution}
    \bra{Q} e^{-HT}\ket{Q }  \underset{T\to \infty}{=} \tilde{ \mathcal{N}} e^{-E_Q T} {=} \tilde{ \mathcal{N}} e^{-\frac{\Delta_Q}{R} T} \ ,
\end{equation}
with $\tilde{ \mathcal{N}}$ a normalization factor. As discussed in \cite{Antipin:2020abu}, the anomalous dimension of $T_Q$ is not affected by the number of Cartan charges \eqref{cartancharges} we fix as long is different from $0$. In other words, one is free to rotate all the non-zero Cartan charges of the $O(N)$ vector model into one single component without loss of generality. This special property is known to apply to the $O(N)$ vector model and $\Delta_Q$ only and fails in more general cases where the distribution of individual Cartan charges affects the physics~\cite{Antipin:2021akb, Antipin:2020rdw, Hellerman:2018sjf}. For the sake of simplicity, we fix only one charge to $Q$. Then the solution of the EOM with the minimal energy is spatially homogeneous and reads \cite{Antipin:2020abu}
%\tad{Chen:$t$ should be changed to $\tau$ which is the cylinder time introduced at the step of Weyl map.}
\begin{equation} \label{SolEOMcubic}
  \begin{cases}
\rho =  f\, \ ,  \ \  \ \chi =- i \mu \tau  \ , \ \ \eta = v \ ,\\
   \varphi_i =  0 \, \qquad i=2,\dots, N/2  \ .
\end{cases}
\end{equation}
where $\varphi_1 =  \frac{1}{\sqrt{2}}\rho e^{i \chi}$ and $\mu$ has the role of the chemical potential associated with the fixed charge. The parameters $f$, $v$, and $\mu$ are fixed by the EOM and the expression for the Noether charge as
\begin{equation} \label{paramm}
    \mu^2-m^2 = g_0 v \ , \qquad \quad  \frac{g_0}{2}f^2 +  \frac{h_0}{2}v^2+ m^2 v = 0 \ , \qquad \quad \frac{Q}{\Omega_{d-1} R^{d-1}}=\mu f^2  \,.
\end{equation}
For convenience, we choose $\ket{Q}$ as
\begin{equation} \label{state}
\ket{Q}  = \int \mathcal{D} \alpha (\vec n)  \,\left\{ \exp \left [ \frac{i Q}{R^{d-1}\Omega_{d-1}} \int d\Omega_{d-1} \, \alpha (\vec n) \right]\right\} \ket{f, 0, \alpha(\vec n), v} \ ,
\end{equation}
where $\vec n$ identifies points on $S^{d-1}$ and $\ket{f, 0, \alpha(\vec n), v}$ is the state with fixed values of the fields $\rho(\vec n) = f$, $\varphi_{i \neq 1}(\vec n) = 0$, $\chi(\vec n) = \alpha (\vec n)$, and $\eta(\vec n) = v$. The term in the brace can be thought of as a wave-functional for the state which fixes one charge to $Q$. Eq.\eqref{state} leads to
\begin{equation} \label{finale}
    \bra{Q}e^{-HT}\ket{ Q} =   \frac{1}{\mathcal{Z}} \int  \mathcal{D} \eta \ \mathcal{D} \varphi \ \mathcal{D} \bar{\varphi} \ e^{- \widehat{\mathcal{S}}} \ ,
    \end{equation}
where $\mathcal{Z}$ is an unimportant normalization constant and
\begin{align}
 \label{O2n_Lagrangian}
\widehat{\mathcal{S}} &= \int^{T/2}_{-T/2} d t \ \int d\Omega_{d-1} \left(   \mathcal{L}_{\rm cyl} + \frac{i Q} {\Omega_{d-1} R^{d-1}}  \ \dot \chi \right) \ . \end{align}
Eq.\eqref{finale} can be computed semiclassically around the solution \eqref{SolEOMcubic}, resulting in
\footnote{As a slight abuse of notation, here we denote the coefficients of the semiclassical expansion as $\Delta_k$, which should not be mistaken for the full scaling dimension $\Delta_Q$.}
\begin{equation}\label{cubicexp}
\Delta_Q =\sum_{k=-1} \frac{1}{Q^k} \Delta_k (\cA) \ , \qquad \cA \equiv Q \eps   \ ,
\end{equation}
which is an expansion in inverse powers of $Q$ at fixed and finite 't Hooft-like coupling $\cA$. This is similar to the semiclassical expansion Eq.\eqref{expansion1}.
%, this is an example of the general phenomenon of \emph{classicalization} of quantum physics in the presence of large quantum numbers \cite{Monin:2016jmo}. \tad{Chen: The method here also applies to small charge, like $Q=1$. This is why the boost works. Thus I suggest removing the sentence.}
The leading order $\Delta_{-1}$ is given by the action $\widehat{\mathcal{S}}$ evaluated on the classical solution. From Eq.\eqref{paramm}, we have
\begin{equation} \label{eqformu}
    R \mu \left[(R \mu)^2 - 4  \right] \left(8 g_0 + h_0  \left[(R \mu)^2 - 4  \right]\right) + \frac{Q g_0^3}{\pi^3} = 0 \ ,
\end{equation}
which, once rewritten at the FP, implicitly defines the chemical potential as a function of the 't Hooft coupling $\cA$. The above equation can be solved numerically, or analytically for small/large values of $\cA$. The first terms of the former expansion reads
\begin{equation} \label{armu}
    R \mu =2 -\frac{g_0^2
  Q}{64 \pi ^3} -\frac{g_0^3 Q^2 (3 g_0+2 h_0)}{16384 \pi ^6}-\frac{g_0^4 Q^3 \left(2 g_0^2 + 2 g_0
   h_0+ h_0^2\right)}{524288 \pi ^9} + \cO\left(Q^4 \right) \ .
\end{equation}
The leading order is the classical energy on the cylinder and reads
\begin{equation}
  Q \frac{\Delta_{-1}}{R} = - \frac{f^2 \mu^2}{2} +\frac{g_0 v f^2}{2} +\frac{h_0 v^3}{6}+\frac{m^2 f^2}{2} +\frac{m^2 v^2}{2}+\frac{Q \mu}{\Omega_{d-1} R^{d-1}} \ .
\end{equation}

Using Eq.\eqref{armu} in the classical energy above and evaluating the result at the FP, we obtain the leading order $\Delta_{-1}(\cA)$ of the semiclassical large-charge expansion. Notice that this classical result resums at once an infinite series of Feynman diagrams. By expanding $\Delta_{-1}$ for small values of $\cA$, we obtain

\begin{align}
\Delta_{-1} &= 2 -\frac{g^2 Q}{128 \pi ^3} -\frac{g^3 Q^2 \left(3 g + 2  h\right)}{49152 \pi ^6} - \frac{ g^4 Q^3 \left(2 g^2 +2 g h + h^2\right)}{2097152 \pi ^9}  -\frac{g^5 Q^4 \left(21 g^3+28 g^2 h+20 g h^2+8 h^3\right)}{1073741824 \pi ^{12}} \nonumber \\ & - \frac{g^6 Q^5 \left(24 g^4+40 g^3 h+36 g^2 h^2+21 g h^3+7 h^4\right)}{51539607552
   \pi ^{15}} -\frac{3 g^7 Q^6 \left(143 g^5+286 g^4 h+308 g^3 h^2+224 g^2 h^3+112 g h^4+32
   h^5\right)}{35184372088832 \pi ^{18}} \nonumber \\ & - \frac{g^8 Q^7 \left(192 g^6+448 g^5 h+560 g^4 h^2+480 g^3 h^3+300 g^2 h^4+132 g
   h^5+33 h^6\right)}{562949953421312 \pi ^{21}} \nonumber \\ & - \frac{11 g^9 Q^8 \left(12597 g^7+33592 g^6 h+47736 g^5 h^2+46800 g^4 h^3+34320
   g^3 h^4+19008 g^2 h^5+7488 g h^6+1664 h^7\right)}{13835058055282163712 \pi
   ^{24}}  +\cO\left( \cA^9 \right) \ ,
   \end{align}
Rewriting the above at the FP
\eqref{gtoxy} gives

\begin{align} \label{resultconcompa}
Q \Delta_{-1}
   &= 2 Q \  - \frac{\eps Q^2}{N}\left(3 +\frac{132}{N} +\frac{5808}{N^2} +\dots \right) - \frac{ Q^3 \eps^2}{N^2}\left(45+\frac{9000}{N}+\frac{3043440}{N^2}+ \dots\right) \nonumber \\ & - \frac{Q^4 \eps^3}{N^3} \left(1350+\frac{495720}{N}+\frac{223974720}{N^2}+ \dots\right) -\frac{Q^5 \epsilon^4}{N^4} \left(\frac{213597}{4}+\frac{28653588}{N}+\frac{15700511880}{N^2}+ \dots\right) \nonumber \\ & - \frac{Q^6 \epsilon ^5}{N^5} \left(2457216 +\frac{1736458560}{N}+\frac{1109489011200}{N^2}+ \dots\right) \nonumber \\ & - \frac{Q^7 \epsilon ^6}{N^6}\left(\frac{995773905}{8}+\frac{109168708635}{N}+\frac{79449296874570}{N^2}+ \dots\right) \nonumber \\ & -\frac{Q^8 \epsilon ^7}{N^7}\left(6739459200+\frac{7060148282880}{N}+\frac{5757420242165760}{N^2}+ \dots\right) \nonumber \\ & - \frac{Q^9 \epsilon ^8}{N^8}\left(\frac{24526111620285}{64}+\frac{9336756738
   09285}{2 N} +\frac{421344743454254565}{N^2}+ \dots\right) + \cO \left( Q^{10} \epsilon ^9\right) \ .
\end{align}
%  \begin{align} \label{resultconcompa}
% Q \Delta_{-1}
%   &= 2 Q \  - \frac{\eps Q^2}{N}\left(3 +\frac{132}{N} +\frac{5808}{N^2} -\frac{1861248}{N^3}+ \cO\left(\frac{1}{N^4}\right)\right) - \frac{ Q^3 \eps^2}{N^2}\left(45+\frac{9000}{N}+\frac{3043440}{N^2}+ \frac{1627608960}{N^3}+ \cO\left(\frac{1}{N^4}\right)\right) \nonumber \\ & - \frac{Q^4 \eps^3}{N^3} \left(1350+\frac{495720}{N}+\frac{223974720}{N^2}+\frac{132387350400}{N^3}+ \cO\left(\frac{1}{N^4}\right)\right)\nonumber \\ &   -\frac{Q^5 \epsilon^4}{N^4} \left(\frac{213597}{4}+\frac{28653588}{N}+\frac{15700511880}{N^2}+\frac{10072777991040}{N^3}+ \cO\left(\frac{1}{N^4}\right)\right) \nonumber \\ & - \frac{Q^6 \epsilon ^5}{N^5} \left(2457216 +\frac{1736458560}{N}+\frac{1109489011200}{N^2}+\frac{768192623815680}{N^3}+ \cO\left(\frac{1}{N^4}\right)\right) \nonumber \\ & - \frac{Q^7 \epsilon ^6}{N^6}\left(\frac{995773905}{8}+\frac{109168708635}{N}+\frac{79449296874570}{N^2}+\frac{59121226858604400}{N^3}+ \cO\left(\frac{1}{N^4}\right)\right) \nonumber \\ & -\frac{Q^8 \epsilon ^7}{N^7}\left(6739459200+\frac{7060148282880}{N}+\frac{5757420242165760}{N^2}+\frac{4587175227804180480}{N^3}+ \cO\left(\frac{1}{N^4}\right)\right) \nonumber \\ & - \frac{Q^9 \epsilon ^8}{N^8}\left(\frac{24526111620285}{64}+\frac{9336756738
%   09285}{2 N} +\frac{421344743454254565}{N^2}+\frac{358159556241651859200}{N^3}+ \cO\left(\frac{1}{N^4}\right)\right) \nonumber \\ &+ \cO \left( Q^{10} \epsilon ^9\right) \ .
% \end{align}
The above remarkably reproduces the $\alpha_j$ coefficients in Eq.\eqref{alfas} for the scaling dimension in the quartic $O(N)$ model. Notice that $\Tilde{\Delta}_{-1} (J)$ in \eqref{compa} and $\Delta_{-1} (\cA)$ in \eqref{resultconcompa} are the leading order in two distinct expansion schemes
denoted respectively as $\sum_{k=-1}\frac{\Delta_k\left(Q/N\right)}{N^k}$ and $\sum_{k=-1}\frac{\Delta_k\left(Q\epsilon\right)}{Q^k}$. However, since at LO in $1/N$ the FP \eqref{gtoxy} is $\cO(\sqrt{\eps})$ exact, all (and only) the terms scaling as $Q\left(\frac{Q \eps}{N}\right)^j$ appear at the LO of both expansions and can be compared. We can thus check terms up to arbitrarily high orders in the conventional loop expansion. Furthermore, we can compare also the term $-132\frac{\eps Q^2}{N^2}$ which is not contained in \eqref{compa} but appears in the diagrammatic result \eqref{1loopfull}. Assuming the validity of the dual description, $\Delta_{-1}$ represents a new result for $\Delta_Q$ in the quartic theory. All the terms scaling as $\frac{Q}{N} \left(\frac{Q \eps}{N}\right)^j$ are contained in the NLO $\tilde{\Delta}_0$ of the semiclassical expansion \eqref{expansion1} and can be used to check future computations of $\tilde{\Delta}_0$ in the quartic theory.

We now move to consider the expansion of $\Delta_{-1}$ for large 't Hooft coupling $\cA$. One needs to select a root of the quintic equation \eqref{eqformu}. This correspond to choosing a root of $ (-1)^{1/5}$. In general, we have
\begin{align} \label{agrree}
   Q \Delta_{-1} & =
  - x \frac{5N}{3}(2\eps)^{1/5} J^{6/5}\left(1+ \cO\left(\frac{1}{N}\right) \right)- \frac{1}{x}\frac{5N}{6}(2\eps)^{-1/5} J^{4/5}\left(1+ \cO\left(\frac{1}{N}\right) \right)\nonumber \\ & + \frac{1}{x^3}\frac{N}{9}(2\eps)^{-3/5} J^{2/5}\left(1+ \cO\left(\frac{1}{N}\right) \right)+\cO \left(J^0 \right)
\end{align}
where the five solutions are parametrized by $x=\left\{1, e^{\pm \frac{4 i \pi}{5}}, e^{\pm \frac{2 i \pi}{5}}\right\}$,
and we have rewritten the result in terms of $J\equiv Q/N$ to compare with the quartic model result \eqref{leadinglarge}. As shown in the Appendix~\ref{selectcp}, the physical (complex conjugate) solutions satisfies the criteria $\rm{Re}[\Delta_{-1}]>0$, which fixes $x= e^{\pm \frac{4 i \pi}{5}}$. This solution matches the quartic result Eq.\eqref{leadinglarge}.

\subsection{Next-to-leading order}

We move to compute the leading quantum correction $\Delta_0$ in the semiclassical expansion, which is given by the functional determinant of the fluctuation around the classical solution \eqref{SolEOMcubic}. First, we note that fixing one charge induces the symmetry breaking pattern below \cite{Antipin:2020abu}
\begin{equation}
	SO(d+1,1) \times O(N) \underset{\text{Explicit}}{\to} SO(d)\times D \times O(N-2) \times U(1) \underset{\text{Spontaneous}}{\to} SO(d)\times D' \times O(N-2)\ ,
\end{equation}
where $D' = D +\mu Q $ with $D$ the generator of the time translations on the cylinder. This symmetry breaking pattern defines a \emph{conformal generalized superfluid} state of matter \cite{Monin:2016jmo, Nicolis:2015sra} which occurs naturally in CFT at fixed-charge. The spontaneous part of the symmetry breaking results in one relativistic Goldstone boson (the so-called conformal phonon), which at large $\mu$ propagates at the speed of sound $c = \sqrt{\frac{1}{d-1}}$ dictated by tracelesness of the energy-momentum tensor. Furthermore, fixing only one charge we are left with $N-2$ "spectator" massive states, with gap $\mu$ and dispersion relation given by \cite{Antipin:2020abu, Alvarez-Gaume:2016vff}
%\tad{Chen: Explain that on the cylinder $p$ is to be understood as $J_\ell$ eventually.}
\begin{equation} \label{dispsect}
    \omega_{*} = \sqrt{p^2 + \mu^2} \ ,
\end{equation}
with $p$ the momentum which is quantized on the cylinder.

For the remaining d.o.f., we expand the fluctuations as follows:
\begin{equation} \label{flutt}
  \begin{cases}
\rho=  f + r(x) \, \ , \ \ \chi =- i \mu \tau + \frac{\pi(x)}{f} \ ,\\
\eta = v + \Tilde{\eta}(x) \ .
\end{cases}
\end{equation}
The quadratic Lagrangian for these three modes reads
\begin{equation}
    \widehat{\mathcal{L}}^{(2)}= \frac{1}{2} (\partial r)^2+\frac{1}{2}  (\partial \Tilde{\eta})^2 +  \frac{1}{2}  (\partial \pi)^2 - 2 i \mu r \dot \pi
    + g_0 f \Tilde{\eta} r  + \frac{h_0}{2} v \Tilde{\eta}^2 + \frac{m^2}{2} \Tilde{\eta}^2+\frac{m^2}{2} r^2 \ .
\end{equation}
The dispersion relations can be computed in the momentum space by considering the inverse propagator $\mathcal{P}^{-1}(p)$, which is defined by the quadratic action as
\begin{equation}
 \widehat{\mathcal{S}}^{(2)} = \int \frac{d^d p}{(2 \pi)^d}[r(-p) \,  \ \pi(-p) \,  \ \Tilde{\eta}(-p) ]\, \mathcal{P}^{-1}(p) \begin{bmatrix} r(p) \\ \pi(p) \\ \Tilde{\eta}(p) \end{bmatrix} \ .
\end{equation}

Then the dispersion relations are the positive energy solutions of $\text{det } \mathcal{P}^{-1}(p) = 0$, where
\begin{equation}
  \mathcal{P}^{-1}(p) =  \left( \begin{array}{ccc}
       \frac{1}{2}\left(\omega^2 -p^2 \right)  & i \omega \mu & A \\
      -i \omega \mu   &   \frac{1}{2}\left(\omega^2 -p^2 \right)  & 0 \\
      A & 0 &  \frac{1}{2}\left(\omega^2 -p^2 \right) - B
    \end{array}\right) \ ,
\end{equation}
with
\begin{equation}
    A =\frac{1}{2} \sqrt{\frac{\left( m^2 -\mu^2\right) \left[2 g_0 m^2 +h_0 \left( \mu^2 -m^2\right)\right] }{g_0}} \ , \qquad  \quad B =\frac{1}{2} \left(m^2 + \frac{h_0}{g_0}  \left( \mu^2 -m^2\right)\right) \ .
\end{equation}
It is easy to check that one of the dispersion relations describes the conformal phonon with speed $c =\frac{1}{\sqrt{5}}$ for large $\mu$ and $d\rightarrow 6$.
\vskip 1em

Clearly, at large $N$ the $N-2$ spectator fields provide the leading $N$ contribution to $\Delta_0$, which we denote as $\Delta_0^{(N)}$. Since our goal is to compare with large $N$ results in the quartic model, we start by computing $\Delta_0^{(N)}$. The computation of the functional determinant associated with the spectator fields is given in App.\ref{rino}, together with the details on its renormalization. The final result reads
\begin{align} \label{finalNLO}
\Delta_0^{(N)}= & N\frac{  25 R^6 \mu ^6-130 R^4 \mu ^4-640 R^2 \mu^2 +2304 R \mu-1568
   }{4608}
+\frac{N}{2} \sum_{\ell=1}^\infty\sigma^{(N)}(\ell)\ ,
\end{align}
where the sum over $\ell$ converges and $\sigma(\ell)$ is given by
%\tad{Chen: There is one $l$ in the following equation which should be $\ell$.}
\begin{align}
 \sigma^{(N)}(\ell) &=  \frac{1}{192} \big(16 (\ell+1) (\ell+2)^2 (\ell+3) \sqrt{R^2 \mu ^2+\ell (\ell+4)}-\frac{1}{\ell} \left[R^6\mu ^6+32
   \left(R^2 \mu ^2-1\right) \right. \nonumber \\ &\left. -2 (\ell (\ell+2)+5)R^4 \mu ^4+8 \ell (\ell+2) (\ell (\ell+4)+5)R^2 \mu ^2+16 \ell
   (\ell+2)^3 (\ell (\ell+4)+1) \right]\big) \ .
\end{align}
The sum over $\ell$ can be computed numerically or analytically for small/large values of the 't Hooft coupling. In order to compare with the large $N$ results of Sec.\ref{I}, we note that all the terms scaling as $N \left(\frac{\eps Q}{N}\right)^j$ in $\Delta_0$ receive contribution only form the spectator fields, i.e they can be read off from $\Delta_0^{(N)}$. Furthermore, they have exactly the right scaling to match the $\beta_j$ coefficients in Eq.\eqref{betas}. We have
\begin{align} \label{lista}
     \Delta_0 &= -Q \eps \left[\frac{1}{2}   +\cO\left(\frac{1}{N} \right)  \right] + \frac{( Q \eps)^2}{N} \left[ \frac{7}{4} +\cO\left(\frac{1}{N} \right) \right] + \frac{( Q \eps)^3}{N^2}  \left[\frac{3}{4} (48 \zeta (3)+31) +\cO\left(\frac{1}{N} \right)\right] \nonumber \\ &+ \frac{( Q \eps)^4}{N^3} \left[ \frac{27}{2} (128 \zeta (3)+40 \zeta (5)+41) +\cO\left(\frac{1}{N} \right) \right]  +\frac{(Q \eps)^5}{N^4} \left[\frac{81}{16} (18208 \zeta (3)+7168 \zeta (5)+1792 \zeta (7)+3117)  +\cO\left(\frac{1}{N} \right) \right] \nonumber \\ & + \frac{( Q \eps)^6}{N^5} \left[ 648 (8202 \zeta (3)+3510 \zeta (5)+1218 \zeta (7)+252 \zeta (9)+677)+\cO\left(\frac{1}{N} \right) \right]  \nonumber \\ &+ \frac{( Q \eps)^7}{N^6} \Bigg[  \frac{2187}{32} (4727888 \zeta (3)+2109440 \zeta (5)+836864 \zeta
   (7)+256000 \zeta (9)+45056 \zeta (11)   +110211) +\cO\left(\frac{1}{N} \right) \Bigg] \nonumber \\ &+ \cO\left((Q \eps)^8\right) \ ,
\end{align}
in remarkable agreement with the values listed in Eq.\eqref{betas} for the quartic model.

The subleading $1/N$ orders in every square bracket in the equation above receive contributions also from the fluctuation in Eq.\eqref{flutt}. This can be computed by the same procedure used for the spectator fields and outlined in App.\ref{rino}, but with two differences. First, now one cannot truncate the expressions to the leading $1/N$ order. Second, since the dispersion relations are more involved, the fluctuation determinant has to be regularized and evaluated numerically at fixed values of $N$, $Q$, and $\eps$. Unfortunately, this fact obscures the comparison with the results in the quartic model. We, therefore, limit ourselves to the numerical calculation of the coefficient of the leading $\eps$ term in $\Delta_0$. We extract it by computing numerically $\Delta_0-\Delta_0^{(N)}$ at small values of $g$ and fitting the result to the functional form $(\Delta_0-\Delta_0^{(N)}) \approx C Q g^2$, which follows from Eq.\eqref{armu}. The fit gives $C = 0.0020997(3) \approx \frac{25}{384 \pi^3}$. Considering the result at the FP \eqref{gtoxy} and neglecting the numerical error, we obtain
\begin{equation}
    \Delta_0 = Q \eps \left(-\frac{1}{2}+\frac{4}{N}+\frac{176}{N^2}+\frac{360544}{N^3}+  \cO \left(\frac{1}{N^4} \right)\right) + \cO \left(Q^2 \eps^2 \right)
\end{equation}
It is easy to check that the first three terms match the quartic result Eq.\eqref{1loopfull}. Finally, we numerically estimate the term of order $\frac{\eps^2 Q^2}{N^2}$, which receives contributions from both $\Delta_{-1}$\footnote{In this case, one needs to consider the values of the FP coupling $g^*$ to $2$-loops.} and $\Delta_0$. We have
\begin{equation}
\frac{Q^2 \eps^2}{N^2}\left(\frac{-2219}{2}+1382(2)+155 \right)= \frac{Q^2 \eps^2}{ N^2} \frac{855(4)}{2} \,
\end{equation}
again in agreement with Eq.\eqref{1loopfull}. The three terms in brackets come, respectively, from the $N-2$ spectator fields, the remaining three d.o.f, and $\Delta_{-1}$.
To summarize the comparison with Eq.\eqref{1loopfull}, we add $Q\Delta_{-1}$ to $\Delta_0$ and rewrite our findings as
\begin{align} \label{match3}
   \Delta_Q &= 2 Q - \frac{\eps}{2}Q+ \frac{1}{N} \left[ \left( - 3Q^2+ 4.000(3) Q  \right)\eps +\left( \frac{7}{4}Q^2 -\frac{8}{3}Q \right)\epsilon^2+\cO \left(\eps^3 \right)  \right]+ \frac{1}{N^2}\bigg[ \left( - 132Q^2+ 176.0(1) Q  \right)\eps  \nonumber \\ &  -\left( 45 Q^3 - \frac{855(4)}{2}+ \cO \left(Q \right) \right)\epsilon^2+\cO \left(\eps^3 \right)  \bigg] + \cO\left(\frac{1}{N^3}\right) \ .
\end{align}
The term of order $\frac{Q \eps^2}{N}$ has been estimated by requiring consistency with the known anomalous dimension of $\phi_a$, which reads $\Delta_{\phi_a} = 2 -\frac{\eps}{2} + \frac{1}{N}\left(\eps - \frac{11}{12}\eps^2 + \dots \right) + \cO\left( \frac{1}{N^2}\right)$ and has to stem from Eq.\eqref{match3} when $Q=1$.

All the checks of the duality between the critical cubic and quartic theories are summarized in Table ~\ref{tab:summary}.
%Comment: \usepackage{makecell} at the beginning must be included.
\begin{table}[t]
%\hspace{-2cm}
\begin{center}%\footnotesize
\begin{tabular}{|c|c|c|c|c|c|c|c|c|}
\hline
& &\multicolumn{3}{c|}{O(N) Cubic Theory ($d=6-\epsilon$)} &\multicolumn{4}{c|}{O(N) Quartic Theory ($4<d<6$)}\\ \hline
Operator & Term compared & Eq.\& Ref & \makecell[c]{Expansion\\scheme} & Order & Eq. & \makecell[c]{Expansion\\scheme} & Order & Ref. \\
\hline
$T_Q$ & $Q^{k+1}\left(\frac{\epsilon}{N}\right)^k,k\geq 1$ &  \eqref{resultconcompa}
& $\sum\limits_{k=-1}^\infty\frac{\Delta_k(Q\epsilon,N)}{Q^k}$ & LO & \eqref{alfas}
& $\sum\limits_{k=-1}^\infty\frac{\Delta_k(Q/N,d)}{N^k}$ & LO & \cite{Giombi:2020enj} \\
\hline
$T_Q$ & $(Q\epsilon)^k\left(\frac{1}{N}\right)^{k-1},k\geq 1$ & \eqref{lista}
& $\sum\limits_{k=-1}^\infty\frac{\Delta_k(Q\epsilon,N)}{Q^k}$ & NLO & \eqref{betas}
& $\sum\limits_{k=-1}^\infty\frac{\Delta_k(Q/N,d)}{N^k}$ & LO & \cite{Giombi:2020enj} \\
\hline
$T_Q$ & $\frac{4Q\epsilon}{N}$ & \eqref{match3}
& $\sum\limits_{k=-1}^\infty\frac{\Delta_k(Q\epsilon,N)}{Q^k}$ & NLO & \eqref{1loopfull}
& $\sum\limits_{k=-1}^\infty\frac{\Delta_k(Q/N,d)}{N^k}$ & LO & \cite{Giombi:2020enj}  \\
\hline
$T_Q$ & $-\frac{8}{3}\frac{Q\epsilon^2}{N}$ & \eqref{match3}
& $\sum\limits_{k=-1}^\infty\frac{\Delta_k(Q\epsilon,N)}{Q^k}$ & \makecell[c]{NNLO} & \eqref{1loopfull}
& $\sum\limits_{k=-1}^\infty\frac{\Delta_k(Q/N,d)}{N^k}$ & LO & \cite{Giombi:2020enj} \\
\hline
$T_Q$ & $-\frac{132Q^2\epsilon}{N^2}$ & \eqref{match3}
& $\sum\limits_{k=-1}^\infty\frac{\Delta_k(Q\epsilon,N)}{Q^k}$ & LO & \eqref{1loopfull}
& $\sum\limits_{k=0}^\infty\frac{\Delta_k(Q,d)}{N^k}$ & NLO & \cite{Derkachov:1997ch} \\
\hline
$T_Q$ & $\frac{176Q\epsilon}{N^2}$ & \eqref{match3}
& $\sum\limits_{k=-1}^\infty\frac{\Delta_k(Q\epsilon,N)}{Q^k}$ & NLO & \eqref{1loopfull}
& $\sum\limits_{k=0}^\infty\frac{\Delta_k(Q,d)}{N^k}$ & NLO & \cite{Derkachov:1997ch} \\
\hline
$T_Q$ & $\frac{855Q^2\epsilon^2}{2N^2}$ & \eqref{match3}
& $\sum\limits_{k=-1}^\infty\frac{\Delta_k(Q\epsilon,N)}{Q^k}$ & NLO & \eqref{1loopfull}
& $\sum\limits_{k=0}^\infty\frac{\Delta_k(Q,d)}{N^k}$ & NLO & \cite{Derkachov:1997ch}\\
\hline
$T_Q$ & \makecell[c]{$N\epsilon^{\frac{1-2k}{5}}\left(\frac{Q}{N}\right)^{\frac{6-2k}{5}}$\\$k=0,1,2$} & \eqref{agrree}
& $\sum\limits_{k=-1}^\infty\frac{\Delta_k(Q\epsilon,N)}{Q^k}$ & LO & \eqref{leadinglarge}
& $\sum\limits_{k=-1}^\infty\frac{\Delta_k(Q/N,d)}{N^k}$ & LO & \cite{Giombi:2020enj}\\
\hline \hline
$\phi$ & $\sum\limits_{k=1,j=1}^{k=5,j=3}\frac{\epsilon^k a_{kj}}{N^j}$ & \cite{Fei:2014xta}\cite{Fei:2014yja}\cite{Kompaniets:2021hwg}
& $\sum\limits_{k=0}^\infty \Delta_k(N)\epsilon^k$ & $\rm{N^4LO}$  & NA
& $\sum\limits_{k=0}^\infty\frac{\Delta_k(Q,d)}{N^k}$ & NNLO & \cite{Fei:2014xta}\cite{Fei:2014yja}\cite{Vasiliev:1982dc} \\
\hline
$\sigma$ & $\sum\limits_{k=1,j=1}^{k=3,j=2}\frac{\epsilon^k b_{kj}}{N^j}$ & \cite{Fei:2014xta}\cite{Fei:2014yja}\cite{Kompaniets:2021hwg}
& $\sum\limits_{k=0}^\infty \Delta_k(N)\epsilon^k$ & $\rm{NNLO}$  & NA
& $\sum\limits_{k=0}^\infty\frac{\Delta_k(Q,d)}{N^k}$ & NLO & \cite{Fei:2014xta}\cite{Fei:2014yja}\cite{Vasiliev:1981dg}\\
\hline
\makecell[c]{$(\phi_i\phi_i)^2$\\$\sigma^2;\phi_i\phi_i$}
& $\sum\limits_{j=1}^{j=2}\frac{\epsilon c_j}{N^j}$  &  \cite{Fei:2014xta}\cite{Fei:2014yja}
& $\sum\limits_{k=0}^\infty \Delta_k(N)\epsilon^k$ & NLO & NA
& $\sum\limits_{k=0}^\infty\frac{\Delta_k(Q,d)}{N^k}$ & NLO & \cite{Fei:2014xta}\cite{Fei:2014yja}\cite{Broadhurst:1996ur} \\
\hline
\makecell[c]{$\sigma^3$\\$\sigma^3;\sigma\phi\phi$} & $-\frac{420\epsilon}{N}$ &  \cite{Fei:2014xta}\cite{Fei:2014yja}
& $\sum\limits_{k=0}^\infty \Delta_k(N)\epsilon^k$ & LO & NA
& $\sum\limits_{k=0}^\infty\frac{\Delta_k(Q,d)}{N^k}$ & LO & \cite{Fei:2014xta}\cite{Fei:2014yja}\cite{Lang:1992zw}\\
\hline
\end{tabular}
\end{center}
\caption{\label{tab:summary}
Summary of the
% terms of the operator scaling dimensions for the equivalence
duality between $O(N)$ cubic theory and quartic theory. In each line, we show a term of the scaling dimensions which matches between the $O(N)$ cubic and quartic theories together with the expansion scheme according to which it has been computed.
All the comparisons made in this work have been summarized in the upper part of the table (above the two-line dividing line). The lower part of the table covers the existing results in the literature where NA denotes not applicable. In the last two lines of the table, $\left(\phi_i\phi_i\right)^2$ (similarly $\sigma^3$) operator in the quartic theory matches to the primary operator from mixing of the $\sigma^2$ and $\phi_i\phi_i$ (similarly $\sigma^3,\,\sigma\phi\phi$) in the cubic theory.
}
\end{table}

\section{Complex anomalous dimensions} \label{complex}
\begin{figure}
\begin{center}
\includegraphics[width=0.45\textwidth]{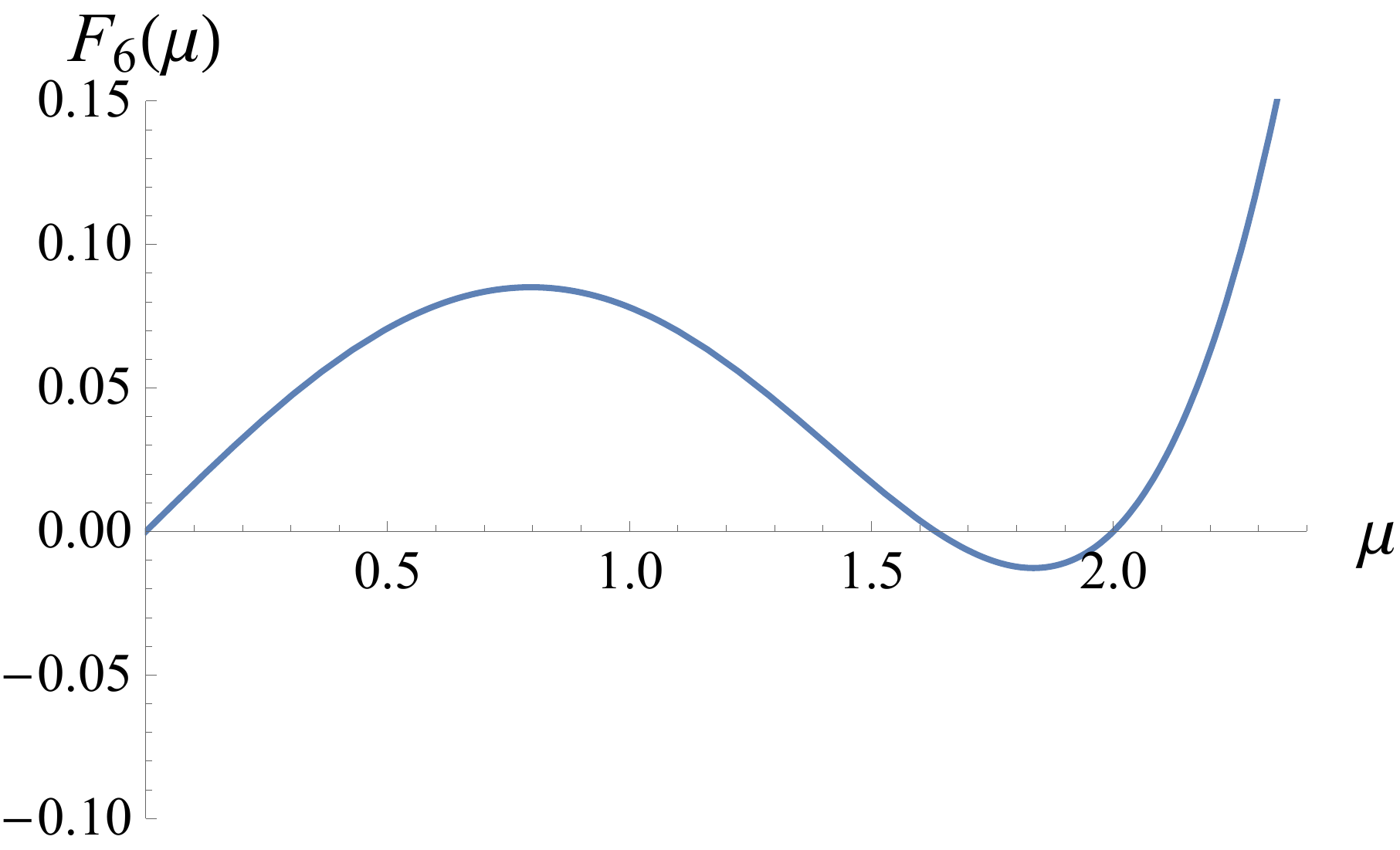} $\qquad$ \includegraphics[width=0.45\textwidth]{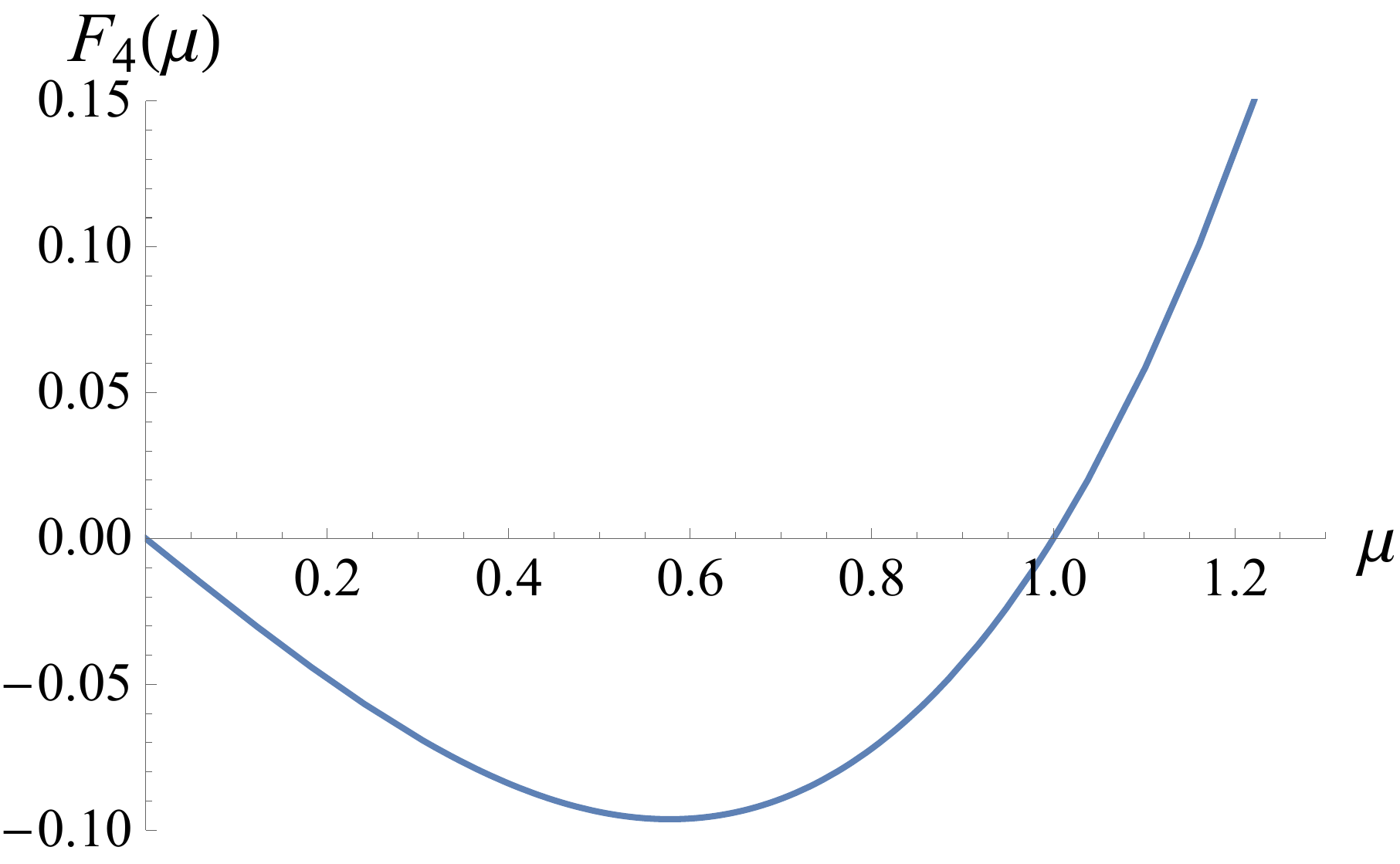}
\caption{\label{plotFaf}$F_6(\mu)$ (\emph{Left}) and $F_4(\mu)$ (\emph{Right}) as a function of $\mu$. The value of these two functions at their minimum at positive values of $\mu$ determines $Q_c$ in $d=6-\eps$ and $d=4+\eps$, respectively.}
\end{center}
\end{figure}

As for the quartic model in $4<d<6$, our result \eqref{agrree} reveals the existence of a critical value of the charge $Q_c$ above which the scaling dimensions are complex. Here we analytically estimate $Q_c$ in $d=6-\eps$ in the cubic model and in $d=4+\eps$ in the quartic model to the leading order in the $\eps$-expansion. In $d=6-\eps$ dimensions, the imaginary part occurs first in the chemical potential $\mu$, which is given implicitly by Eq.\eqref{eqformu}. Let us fix $R=1$ and rewrite Eq.\eqref{eqformu} at the FP \eqref{gtoxy}. Truncating the FP values at the leading order in $1/N$, Eq.\eqref{eqformu} can be transformed into
\begin{equation} \label{rewrite}
  F_6(\mu) \equiv \frac{1}{192} \mu \left(3 \mu^4-20 \mu^2+32\right)=- \frac{Q \eps}{N} +\cO\left(\frac{1}{N^2}\right)  \ .
\end{equation}
The plot of $F_6(\mu)$ in the physical region $\mu>0$ is shown in Fig.\ref{plotFaf};
we note that only a limited range of values of the product $ \frac{Q\epsilon}{N} >0 $ allows for a real and positive chemical potential. As in the $O(N)$ case, above $Q_c \equiv Q_c (\eps, N)$ there are no physical solutions to the saddle-point equations, and the scaling dimensions acquire an imaginary part. $Q_c$ is determined by the value of $F_6(\mu)$ at its minimum
\begin{equation}
    Q_c  = \frac{N}{90 \eps} \left(-9 + \sqrt{105}\right) \sqrt{\frac{1}{30} \left(15 + \sqrt{105}\right)}  \ .
\end{equation}
% In Fig.\ref{plotmudia} we show the real part of the chemical potential as a function of the product $\frac{Q \eps}{N}$.
At $Q=Q_c$, the chemical potential is non-analytic. Notice that, since the $1$-loop FP is real only when $N>1038$, in this range of values of $N$, the inclusion of the subleading $1/N$ orders in Eq.\eqref{rewrite} results solely in small corrections to $Q_c$.
% \begin{figure}
% \begin{center}
%  \includegraphics[width=0.45\textwidth]{plotrealmu.pdf}
% \caption{$Re(\mu)$ as a function of $\frac{Q \eps}{N}$. The spike is allocated in $Q_c = \frac{N}{90 \eps} \left(-9 + \sqrt{105}\right) \sqrt{\frac{1}{30} \left(15 + \sqrt{105}\right)}$.}
% \label{plotmudia}
% \end{center}
% \end{figure}

For the sake of completeness, we study the quartic $O(N)$ model in $d=4+\eps$, which is obtained by continuing our results \cite{Antipin:2020abu} in $d=4-\eps$, to negative $\eps$. There, we studied $\Delta_Q$ in the double scaling limit $\eps \to 0$, $Q \to \infty$ with $Q \eps$ fixed. This results in a semiclassical expansion analogous to Eq.\eqref{cubicexp}. In that case, we found the chemical potential of the system as the solution of a cubic equation
\begin{equation} \label{mu4plus}
\mu^3 -  \mu = \frac{4}{3} Q g^* \ ,
\end{equation}
with $g^* = g^*(\eps)$ the fixed point coupling.
The physical solution, which is real below $d=4$ and matches perturbation theory for small $Q g^*$, reads
\begin{equation} \label{nonso}
  R \mu = \frac{3^\frac{1}{3}+\left(6 g^* Q + \sqrt{-3+36 (g^* Q)^2}\right)^\frac{2}{3}}{3^\frac{2}{3}\left(6 g^* Q + \sqrt{-3+36 (g^* Q)^2}\right)^\frac{1}{3}} \ .
\end{equation}
As before, to study the appearance of complex scaling dimensions, we rewrite Eq.\eqref{mu4plus} at the FP $g^*(\eps) =- \frac{3}{N+8} \eps$  as
\begin{equation} \label{cacciaF4}
    F_4(\mu) \equiv \frac{1}{4} \left(\mu ^3-\mu \right)=-\frac{Q \eps}{N+8} + \cO\left(\eps^2 \right) \ .\end{equation}
The plot of $F_4(\mu)$ for $\mu > 0$ is shown in Fig.\ref{plotFaf}: we have two regimes corresponding to $\eps$ positive and negative. For negative $\eps$ we are in $d<4$, $F_4(\mu)$ is positive and monotonic and there are no complex anomalous dimensions, as expected. For positive $\eps$, there is a minimum in $\mu = \frac{1}{\sqrt{3}}$, and we have
\begin{equation}
Q_c =-(N +8)\frac{F_4\left(\frac{1}{\sqrt{3}}\right)}{\eps} = \frac{N+8}{6 \sqrt{3} \eps} \ ,
\end{equation}
In general, by using Eq.~\eqref{mu4plus} one can obtain $Q_c = - \frac{1}{2 \sqrt{3} g^*(\eps)}$ and study the corrections to $Q_c$ due to higher $\eps$ orders in $g^*(\eps)$.

\begin{figure}
\begin{center}
\includegraphics[width=0.5\textwidth]{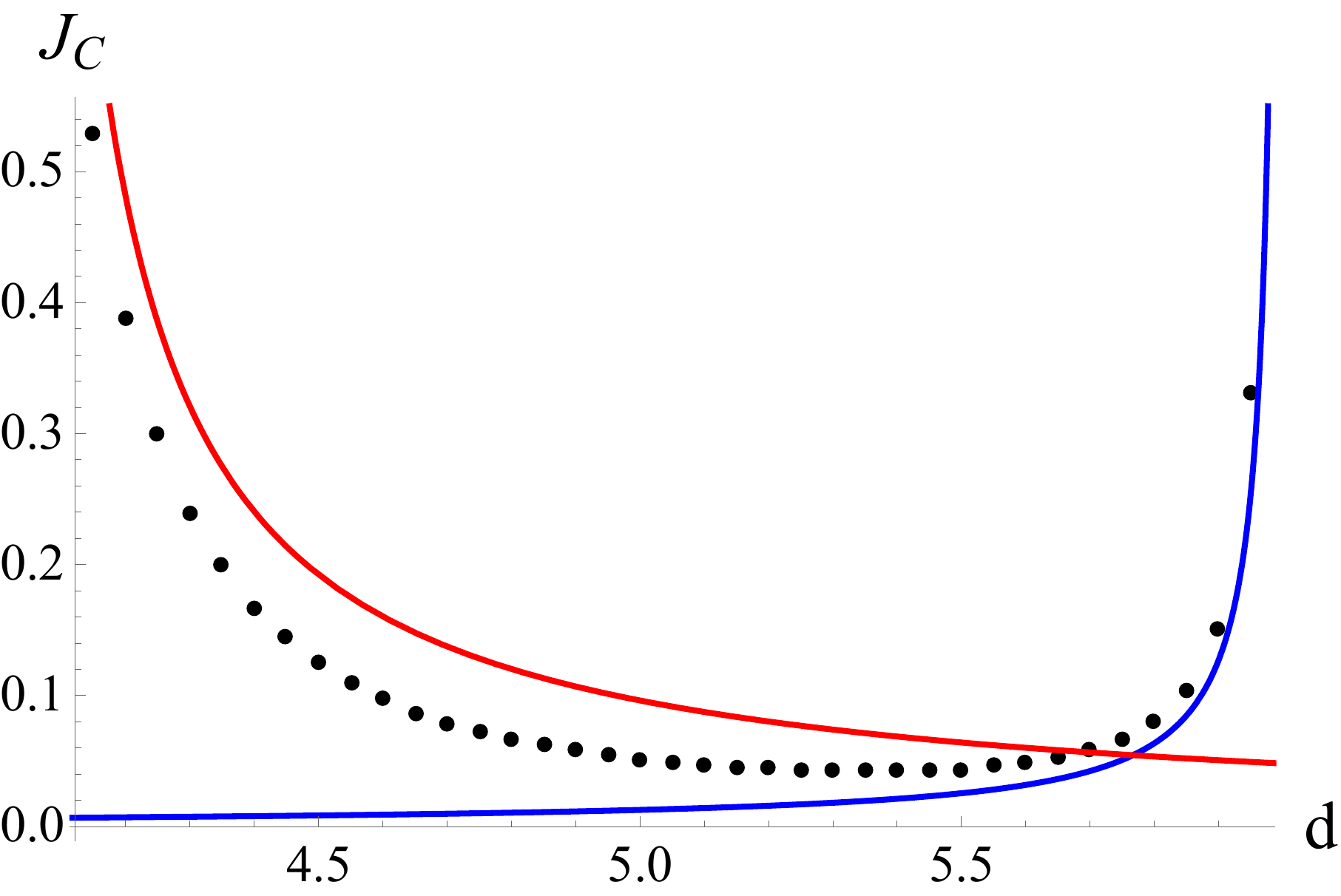}
\caption{\label{Jctot} The behaviour of $J_c$ in $d=4+\eps$ (red curve) and $d=6-\eps$ (blue curve) to leading order in both $1/N$ and  $\eps$. The black dots correspond to the numerical estimation of $J_c$ to leading order in $1/N$ obtained in \cite{Giombi:2020enj}.}
\end{center}
\end{figure}

Finally, to make contact with the numerical estimation of $Q_c$ in $4<d<6$ made in \cite{Giombi:2020enj}, we consider large $N$ and introduce $J_c (d) \equiv Q_c(d)/N$. In Fig.\ref{Jctot}, we show the two tails we found for $J_c$ around $6$ and $4$ dimensions together with the numerical result of \cite{Giombi:2020enj}.
% \footnote{Since at large $N$ the FP \eqref{gtoxy} are $1$-loop exact in $\eps$, the difference with \cite{Giombi:2020enj} does not come from the truncation of the FP value but stems from higher-order terms in $\epsilon$, which are sub-leading in our semiclassical expansion \eqref{cubicexp} but not in the semiclassical expansion \eqref{expansion1}.} .

\section{Discussion} \label{discuss}
We have investigated the large-charge dynamics of the cubic version of the $O(N)$ model for any $N$ just below six dimensions. In this limit, we computed the scaling dimensions of a family of fixed-charge operators at the infrared fixed point of the model to leading and subleading order in the fixed charge expansion but to all orders in the couplings.  The so obtained results allowed us to investigate the conjectured equivalence with the $O(N)$  model featuring quartic interactions at its ultraviolet fixed point. We compared the newly derived information on the scaling dimensions with the known large $N$ results for the  quartic interaction model and shown that they agree.
 Our work therefore strengthens the conjectured equivalence while providing novel information on the finite $N$ physics  coming from our computations within the critical cubic model just below 6 dimensions. Finally, the results presented here could also be of useful in holography since it is believed that $d$-dimensional $O(N)$ CFTs can have a holographic description in terms of Vasiliev higher-spin theories in $AdS_{d+1}$ \cite{Fradkin:1987ks, Klebanov:2002ja, Giombi:2012ms}.

\section*{Acknowledgements}
The work of O.A. and J.B. is partially supported by the Croatian Science Foundation project number 4418.
C.Z. is supported by MIUR under grant number 2017L5W2PT and INFN grant STRONG and thanks the Galileo Galilei Institute for the hospitality.

\appendix

\section{Selecting the chemical potential in the large 't Hooft coupling regime} \label{selectcp}

The main point of this appendix is to show that the values of $x$ in \eqref{agrree} to be chosen are the ones for which the real part of the scaling dimension is positive. Arguing for continuity between large and small charge values compatible with the semiclassical expansion we can use the information at small charge to find a lower positive bound on the real part of the scaling dimension at large charge.

We now move to prove the statement above. In the semiclassical analysis a key equation is the relation between the chemical potential $\mu$ and the charge $Q$, Eq.~\eqref{eqformu}, from which one eliminates $\mu$ and expresses the scaling dimension $\Delta_Q$ solely in terms of $Q$ (and $N,\epsilon$). This equation is however quintic, allowing five solutions for $\mu$, in which one is real, and the other four are complex. In the small 't Hooft coupling regime (i.e. $\frac{Q\epsilon}{N}\ll 1$), one  chooses the  solution that matches the perturbative result for $\Delta_Q$. In the large 't Hooft coupling regime (i.e. $\frac{Q\epsilon}{N}\gg 1$) other methods to select among the solutions are needed. As discussed below Eq.~\eqref{agrree}, this amounts to select the value of $x$ from the set $\left\{1, e^{\pm \frac{4 i \pi}{5}}, e^{\pm \frac{2 i \pi}{5}}\right\}$.

In the large 't Hooft coupling regime, the condition $\frac{Q\epsilon}{N}\gg 1$ implies that the leading contribution
in Eq.~\eqref{agrree} comes from the first term on its right-hand side, which we now write as
\begin{align}
T_1\equiv- x \frac{5N}{3}(2\eps)^{1/5} J^{6/5}
\end{align}
where we ignored the $\mathcal{O}(N^{-1})$ correction. With this expression, only $x=e^{\pm \frac{4 i \pi}{5}}$ leads to a positive real part for $T_1$ (and accordingly for $\Delta_Q$). In the present context, it is however not possible to use unitarity bound to exclude the other choices $x=1,e^{\pm \frac{2 i \pi}{5}}$ which yields a negative real part
for $\Delta_Q$. The reason is that we have not proven the $O(N)$ cubic theory in $d=6-\epsilon$ dimensions is unitary.
In fact, it is likely to be non-unitary for two reasons. First, the Wilson-Fisher FPs associated with fractional dimensions are known to be non-unitary~\cite{Hogervorst:2015akt}. Second, the $O(N)$ quartic theory is known to exhibit complex scaling dimensions in the large charge sector~\cite{Giombi:2020enj}. If the equivalence between the cubic and quartic theories holds, then the cubic theory should also exhibit complex scaling dimensions in the large charge sector. 

Nevertheless,  as we shall see, we are able to set a lower bound on $\rm{Re}\Delta_Q$. We start by noticing that $T_Q$ is actually an irreducible tensor multiplet with components corresponding to weights of the $Q$-index traceless symmetric tensor representation of $O(N)$. (A weight is just a charge configuration in Lie algebraic terms.) Then the bound on $\rm{Re}\Delta_Q$ can be obtained following the reasoning below:
\begin{enumerate}
\item For a charge configuration with its total charge $Q$ in the large 't Hooft coupling regime, the irreducible tensor multiplet associated with $T_Q$ must contain some component operator $O_S$ corresponding to a charge configuration with total charge $Q_S$ in the small 't Hooft coupling regime\footnote{There is a simple analogy with the $SU(2)$ case where the total charge corresponds to the total spin while the charge configuration corresponds to its projection along a given direction.}. (To be proven below.)
\item In the irreducible tensor multiplet associated with $T_Q$, all component operators have the same scaling dimension according to the Wigner-Eckart theorem. ($\rm{Re}\Delta_Q=\rm{Re}\Delta_{O_S}$)
\item Semiclassical computations in the small 't Hooft coupling regime sets a lower bound on the real part of the scaling dimension for all the possible operators associated with the same charge configuration. 
%This means that different irreducible representations $T_Q$ can share the same charge configuration ${Q_S}$. 
%Nevertheless 
%Only one representation $T_{Q_S}$ can have the smallest positive  real scaling dimension $ \rm{Re}\Delta_{Q_S}$.
In fact we have $\rm{Re}\Delta_{Q_S}>0$ where $\Delta_{Q_S}$ is the lowest-lying scaling dimension among the operators with the same charge configuration as $O_S$. Then we have: $\rm{Re}\Delta_{Q}=\rm{Re}\Delta_{O_S}\geq \rm{Re}\Delta_{Q_S}>0$.
\end{enumerate}
Here point 3 is related to the fact that we are computing the matrix element $\bra{Q} e^{-HT}\ket{Q }$ in the limit of
$T\rightarrow +\infty$ which projects out the contribution of the lowest-lying operator with fixed-charge $Q$. When the scaling dimension can be complex, ``lowest-lying" refers to the real part of the scaling dimension. For a charge configuration $Q_S$ in the small 't Hooft coupling regime, conventional perturbation theory can be trusted which indicates the scaling dimension $\Delta_{Q_S}$ associated with the lowest-lying operator in the charge configuration $Q_S$ must be real and positive, i.e. $\Delta_{Q_S}>0$. Therefore the combination of the three points above leads to
$\rm{Re}\Delta_Q>0$ even for $Q$ in the large 't Hooft coupling regime, allowing us to select $x=e^{\pm \frac{4 i \pi}{5}}$.

The point 1 above can be proven using Theorem 10.1 of ref.~\cite{Hall:2015tb}. Suppose we have an irreducible representation $\Gamma$ of a complex semi-simple Lie algebra with the highest weight $\nu$, the theorem then states that an integral element (i.e. integer combination of fundamental weights) $\lambda$ is a weight of $\Gamma$ if and only if the following two conditions are satisfied:
\begin{itemize}
\item $\lambda$ belongs to the convex hull of the Weyl-group orbit of $\nu$.
\item $\nu-\lambda$ can be expressed as an integer combinations of roots.
\end{itemize}
Here the convex hull of a set of vectors $v_1,...,v_M$ is defined to be the set of all vectors of the form $c_1 v_1+c_2 v_2+...+c_M v_M$ where the $c_j$'s are non-negative real numbers satisfying $c_1+c_2+...+c_M=1$.

Intuitively, the convex hull of the Weyl-group orbit of the highest weight $\nu$ of $T_Q$ must
encompass a neighborhood of origin in which one can find some integral element $\nu_S$ that
can be obtained from $\nu$ by subtracting an integer combination of roots. Because $\nu_S$ is
close to the origin, it must be in the small 't Hooft coupling regime. Thus point 1 follows from the theorem.

We can make the above intuitive understanding rigorous by explicitly finding an integral element $\nu_S$ in the small
't Hooft coupling regime which simultaneously satisfies the two conditions of the theorem.

First consider the case of odd $N$, that is $N=2l+1$ with $l$ a positive integer. This corresponds to $B_l$ Lie algebra and $T_Q$ is associated with the highest weight $\nu=Q\Lambda_1$, with $\Lambda_1$ being the first fundamental weight which can be expressed in terms of the positive simple roots $\alpha_1,\alpha_2,...,\alpha_l$ of $B_l$ (c.f. Appendix F of ~\cite{Cornwell:1985xt})
\begin{align}
\Lambda_1=\sum_{p=1}^l \alpha_p
\end{align}
Now note that $T_Q$ is a real representation, $\bar{\nu}\equiv-\nu$ must be a weight of $T_Q$ and therefore be in the convex hull of the Weyl-group orbit of $\nu$ according to the theorem. Thus let us consider
\begin{align}
\nu_S\equiv\frac{Q+1}{2Q}\nu+\frac{Q-1}{2Q}\bar{\nu}
\end{align}
from which we easily find $\nu_S=\Lambda_1$ is an integral element. Also $\nu_S$ belongs to the convex hull $\nu$ and $\bar{\nu}$, and therefore it belongs to\footnote{Using the definition of the convex hull, it is straightforward to prove that if $C$ is the convex hull of $v_1,...,v_M$, then for any $u_1,...,u_K\in C$, the convex hull of $u_1,...,u_K$ must be a subset of $C$.} the convex hull of the Weyl-group orbit of $\nu$. Finally we may easily confirm that $\nu-\nu_S=(Q-1)\sum_{p=1}^l \alpha_p$ which is an integer combination of roots. Therefore $\nu_S$ satisfies all conditions of the theorem. Moreover $\nu_S$ is in the small 't Hooft coupling regime since $\nu_S=\Lambda_1$ is the charge configuration associated with an $O(N)$ vector.

Second, consider the case of even $N$, that is $N=2l$ with $l$ a positive integer. This corresponds to $D_l$ Lie algebra and $T_Q$ is associated with the highest weight $\nu=Q\Lambda_1$, with $\Lambda_1$ being the first fundamental weight which can be expressed in terms of the positive simple roots $\alpha_1,\alpha_2,...,\alpha_l$ of $D_l$ (c.f. Appendix F of ~\cite{Cornwell:1985xt})
\begin{align}
\Lambda_1=\sum_{p=1}^{l-2} \alpha_p+\frac{1}{2}\alpha_{l-1}+\frac{1}{2}\alpha_l
\end{align}
Again note that $T_Q$ is a real representation, $\bar{\nu}\equiv-\nu$ must be a weight of $T_Q$ and therefore be in the convex hull of the Weyl-group orbit of $\nu$. Thus let us consider
\begin{align}
\nu_S &\equiv\frac{Q+1}{2Q}\nu+\frac{Q-1}{2Q}\bar{\nu},\,\,\,\rm{odd}\,Q, \\
\nu_S &\equiv\frac{Q+2}{2Q}\nu+\frac{Q-2}{2Q}\bar{\nu},\,\,\,\rm{even}\,Q,
\end{align}
By simple computation we find
\begin{align}
\nu_S &=\Lambda_1,\,\,\,\rm{odd}\,Q, \\
\nu_S &=2\Lambda_1,\,\,\,\rm{even}\,Q,
\end{align}
and
\begin{align}
\nu-\nu_S &=(Q-1)\sum_{p=1}^{l-2} \alpha_p+\frac{Q-1}{2}\alpha_{l-1}+\frac{Q-1}{2}\alpha_l,\,\,\,\rm{odd}\,Q, \\
\nu-\nu_S &=(Q-2)\sum_{p=1}^{l-2} \alpha_p+\frac{Q-2}{2}\alpha_{l-1}+\frac{Q-2}{2}\alpha_l,\,\,\,\rm{even}\,Q,
\end{align}
Therefore $\nu_S$ obviously satisfies the conditions of the theorem. It is also in the small 't Hooft coupling regime as it corresponds to the charge configuration of an $O(N)$ vector or 2-index traceless symmetric $O(N)$ tensor.
%From these considerations, we then obtain a positive lower bound for $\rm{Re}\Delta_Q$, which determines $x=e^{\pm \frac{4 i \pi}{5}}$.

\section{Computation of $\Delta_0^{(N)}$} \label{rino}

In this appendix, we illustrate the computation of the contribution of the spectator fields $\Delta_0^{(N)}$. Being this a quantum contribution, we will need to renormalize our results. Then, we start by rewriting the expansion \eqref{cubicexp} in its bare and renormalized forms
\begin{equation}
E_Q^{(N)} R = \sum_{k=-1}^{\infty} \frac{1}{Q^k} e_k^{(N)}(g_{0}, h_{0}, Q, d) =\sum_{k=-1}^{\infty} \frac{1}{Q^k} \bar e_k^{(N)}(g, h, Q, d, R M) \ , \label{bareren}
\end{equation}
where $e_j^{(N)}$ and $\bar e_j^{(N)}$ are, respectively, the bare and renormalized coefficients of the expansion and $M$ is the renormalization scale. $e_0^{(N)}$ is determined by the functional determinant of the spectators' fluctuations and can be written in terms of the dispersion relations \eqref{dispsect} as \cite{Badel:2019oxl,Antipin:2020abu}
\begin{equation} \label{masterclass}
e_0^{(N)} (g_{0}, h_{0}, Q, d)=
% \frac{R}{T} \log \frac{\sqrt{\det \widehat{\mathcal{S}}^{(2)}}}{\det\left (- \partial_\tau^2-\Delta_{S^{d-1}} +m^2\right )} =
N \frac{R}{2}\sum_{\ell=0}^\infty n_{\ell}  \omega_* (p^2 =J_\ell) \ ,
\end{equation}
where the factor of $N$ comes from summing over all the spectator modes. The expressions of the eigenvalues of the Laplacian on the sphere $J_\ell$ and their multiplicity $n_\ell$ are given by
\begin{equation}
J_\ell^2=\frac{\ell\left(\ell+d-2\right)}{R^2}\,, \qquad \quad n_{\ell}=\frac{\left(2\ell+d-2\right)\Gamma\left(\ell+d-2\right)}{\Gamma\left(\ell+1\right)\Gamma\left(d-1\right)}\,.
\end{equation}
The sum over $\ell$ in \eqref{masterclass} diverges and needs regularization. The renormalization is carried out at the one-loop level and leading order in $1/N$. Working in MS scheme, bare and renormalized couplings are related by \cite{Machacek:1983tz}
\begin{equation}\label{eq:Z-factors}%
	 g_0 = M^{\eps/2}  \left(g+ \sum_{k=1}^{\infty} \frac{Z_{g,k}(g, h)}{\eps^k}\right)  \ , \qquad \qquad h_0 = M^{\eps/2}  \left(h+ \sum_{k=1}^{\infty} \frac{Z_{h,k}(g, h)}{\eps^k}\right) \ ,
\end{equation}
% where
% \begin{equation}
% Z_g
% 	= 1 + \sum_{k=1}^{\infty} \frac{Z_{g,k}(g, h)}{\eps^k} \ , \qquad \qquad Z_h
% 	= 1 + \sum_{k=1}^{\infty} \frac{Z_{h,k}(g, h)}{\eps^k} \ .
% 	\label{eq:Z-factor-expansion}
% \end{equation}
The beta functions of the couplings are related to $Z_{g,k}$ and $Z_{h,k}$ as
\begin{equation}
	2 \beta_g
	= -\eps g +  g \frac{\partial Z_{g,1}(g ,h)}{\partial g} + h \frac{\partial Z_{g,1}(g, h)}{\partial h} - Z_{g,1}(g ,h) \ , \qquad \qquad 	2 \beta_h
	= -\eps h +  g \frac{\partial Z_{h,1}(g ,h)}{\partial g} + h \frac{\partial Z_{h,1}(g, h)}{\partial h} - Z_{h,1}(g ,h)  .
	\label{eq:beta-def}%
\end{equation}
Then, taking the large $N$ limit in Eq.\eqref{beta1}, one has
\begin{equation} \label{relation}
g_{0} = M^{\eps/2} \left(g + \frac{g^3 N}{768 \pi ^3 \eps} \right) \ , \qquad h_{0} = M^{\eps/2} \left(h +\frac{g^2 N (h-4 g)}{256 \pi ^3 \eps}\right) \ ,
\end{equation}
The renormalization is performed by using Eq.\eqref{relation} into Eq.\eqref{bareren} and expanding every term in powers of couplings. This procedure mixes the bare orders of the expansion. In particular, we have
\begin{equation}
\bar e_0^{(N)}(g, h, Q, d, R M) = e_0^{(N)} (g, h, Q, d)+ f_0^{(N)} (g, h, Q, d, R M) \ ,
\end{equation}
where
\begin{align}
  f_0^{(N)} & = \frac{N \left(\mu ^2 R^2-4\right)^2 \left(\mu ^2 R^2-2\right)}{384 R }\left(\frac{1}{\eps}-\log(M R \sqrt{\pi}) \right)  -\frac{N }{1536 R} \nonumber \\ & \times \bigg(\left(\mu ^2 R^2-4\right) \left((2 \gamma -3) \mu ^4 R^4-12 \gamma  \mu ^2 R^2-2 \mu ^2 R^2+16 \gamma +24\right)\bigg) + \cO(\eps) \ ,
\end{align}
with $\gamma$ is the Euler–Mascheroni constant. We computed $f_0$ by expanding $e_{-1}$ (which is obtained as $\Delta_{-1}$ but working in $d=6-\eps$ instead of $d=6$) in powers of the couplings and retaining the term of order $g_0$ and $h_0$.
The next step is to evaluate this expression at the FP. Being the fixed point couplings expressed as a power series in $\eps$, this step mixes again different orders of the expansion, now the renormalized ones. In particular, in order to include in $\Delta_0^{(N)}$ all the terms with the right scaling, we need to add the expansion of $Q \bar e_{-1}$ to the leading order in $\eps$. After this procedure, the term depending on the renormalization scale $M$ drops, and $\Delta_0^{(N)}$  depends only on $\cA = Q\eps$. We have
\begin{align} \label{deltaxme}
\Delta_0^{(N)} (\cA) &=  N \left\{\underset{\eps\to 0}{
\text{lim}}\left[\frac{R}{2}\sum_{\ell=0}^\infty n_{\ell}
\omega_*(\ell)
+\frac{\left(\mu ^2 R^2-4\right)^2 \left(\mu ^2 R^2-2\right)}{384 R \eps}\right]\right\}_{g, h = g^*(\eps), h^*(\eps)} \ .
\end{align}
The sum over $\ell$ can be regularized as done in \cite{Badel:2019oxl,Antipin:2020abu}. In the regularization procedure, also the $\frac{1}{\eps}$ pole in Eq.\eqref{deltaxme} cancels, and we can consistently take the limit $\eps \to 0$ in Eq.\eqref{deltaxme}, after which we are left with our final result \eqref{finalNLO}, which is finite. Notice that, since the two $1/\eps$ terms come from different orders of the bare expansion, their cancellation can be used as a non-trivial internal check of the correctness of our calculations. We checked numerically that the cancellation of the $1/\eps$ pole occurs also in the renormalization of the full $\Delta_0$ coefficient.

  \end{document}